\documentclass[a4paper,12pt]{article}
\pdfoutput=1

\usepackage{jheppub}

\usepackage{graphicx}
\usepackage{amssymb}
\usepackage{amsmath}
\usepackage{epstopdf}
\usepackage{bm}
\usepackage{color}
\usepackage[english]{babel}
\usepackage[utf8]{inputenc}
\usepackage[T1]{fontenc}
\usepackage[dvipsnames]{xcolor}
\usepackage{color}
\usepackage{bbold}
\usepackage{capt-of}
\usepackage{pstricks}
\usepackage{bm} 
\usepackage{slashed}
\usepackage{booktabs}
\usepackage{tikz}

\long\def\comment#1{ }

\newcommand{\beq}{\begin{eqnarray}}
\newcommand{\eeq}{\end{eqnarray}}
\newcommand{\nn}{\nonumber}

\newcommand{\bea}{\begin{eqnarray}}
\newcommand{\eea}{\end{eqnarray}}
\newcommand{\bean}{\begin{eqnarray*}}
\newcommand{\eean}{\end{eqnarray*}}

\newcommand{\xv}{{\mathbf x}}
\newcommand{\pv}{{\mathbf p}}

\newcommand{\vecnul}{{\mathbf 0}}
\newcommand{\om}{\omega}
\newcommand{\eps}{\epsilon}
\renewcommand{\Im}{\,{\rm Im}\,}
\newcommand{\ttau}{\tilde\tau}
\newcommand{\half}{\frac{1}{2}}
\newcommand{\gmu}{\gamma_\mu}
\newcommand{\tr}{\mbox{tr}\,}
\newcommand{\diag}{\mbox{diag}\,}

\def\bra#1{\langle#1\vert}
\def\ket#1{\vert#1\rangle}

\newcommand{\avg}[1]{\left\langle #1 \right\rangle}

\newcommand{\be}{\begin{eqnarray}}
\newcommand{\ee}{\end{eqnarray}}

\newcommand{\C}{\mathcal{C}}
\renewcommand{\P}{\mathcal{P}}

\newcommand{\Id}{{\mathbb{1}}}

\newcommand{\id}{\mathbb{1}}

\begin{document}

\title{Light baryons below and above the deconfinement transition: medium effects and parity doubling}

\author[1]{Gert Aarts,}
\author[1]{Chris Allton,}
\author[1]{Davide De Boni,}
\author[1]{Simon Hands,}
\author[1,2]{Benjamin J\"ager,}
\author[1]{Chrisanthi Praki}
\author[3,4]{and Jon-Ivar Skullerud}

\affiliation[1]{Department of Physics, College of Science, Swansea University, \\ Swansea SA2 8PP, United Kingdom}
\affiliation[2]{Institute for Theoretical Physics, ETH Z\"urich, CH-8093 Z\"urich, Switzerland}
\affiliation[3]{Department of Theoretical Physics, National University of Ireland Maynooth,\\ Maynooth, County Kildare, Ireland}
\affiliation[4]{School of Mathematics, Trinity College Dublin, Dublin 2, Ireland}

\abstract{
We study what happens to the $N$, $\Delta$ and $\Omega$ baryons in the hadronic gas and the quark-gluon plasma, with particular interest in parity doubling and its emergence as the plasma is heated.
This is done using simulations of lattice QCD, employing the FASTSUM anisotropic $N_f = 2+1$ ensembles, with four temperatures below and four above the deconfinement transition temperature.
Below $T_c$ we find that the positive-parity groundstate masses are largely temperature independent, whereas the negative-parity ones are reduced considerably as the temperature increases.
This may be of interest for heavy-ion phenomenology.
 Close to the transition, the masses are nearly degenerate, in line with the expectation from chiral symmetry restoration.
Above $T_c$ we find a clear signal of parity doubling in all three channels, with the effect of the heavier $s$ quark  visible.
}

\emailAdd{g.aarts@swan.ac.uk}
\emailAdd{c.allton@swan.ac.uk}
\emailAdd{d.de-boni.840671@swan.ac.uk} 
\emailAdd{s.hands@swan.ac.uk}
\emailAdd{bejaeger@itp.phys.ethz.ch}  
\emailAdd{c.s.praki@swan.ac.uk}
\emailAdd{jonivar@thphys.nuim.ie}

\date{\today}


\arxivnumber{1703.09246}

\maketitle


\section{Introduction}
\label{sec:intro}

In the past decade the study of the quark-gluon plasma at the Large Hadron Collider at CERN and the Relativistic Heavy Ion Collider at BNL has matured into a quantitative area of research, in which more detailed questions can be asked and answered -- see e.g.\ Refs.\ \cite{Braun-Munzinger:2015hba,Armesto:2015ioy} and references therein. One topic of interest concerns the changes to the spectrum of QCD, which are expected as hadrons are immersed in a hadronic gas at temperatures below the deconfinement transition, and in the quark-gluon plasma (QGP) at higher temperatures. 
This has been especially important for quarkonium, bound states of a heavy quark and anti-quark, as their melting/survival pattern can act as a thermometer for the temperatures reached in these collisions.
Indeed, both the LHC \cite{Chatrchyan:2012lxa,Abelev:2014nua} and RHIC \cite{Adare:2014hje} have reported clear suppression patterns for bottomonium states at high temperature. Ref.\ \cite{Andronic:2015wma} contains a recent comprehensive review and Ref.\  \cite{Aarts:2016hap} a discussion of open questions.

For light hadrons on the other hand,  the emphasis has been on the statistical properties of the hadrons emerging from the system and on the dilepton spectrum \cite{Braun-Munzinger:2015hba}. Dileptons are predominantly produced by the decay of vector mesons and hence properties of their spectrum provide a connection with chiral symmetry and its restoration at high temperature.
This observation has led to substantial activity on the role of chiral symmetry at finite temperature in the mesonic sector \cite{Rapp:1999ej}.\footnote{Note that throughout this paper chiral symmetry will refer to SU(2)$_{\rm A}$ chiral symmetry, which is spontaneously broken in the vacuum (and explicitly by nonzero quark masses).
}

Due to the nature of the thermal transition in QCD, studies using lattice QCD can provide important nonperturbative insight. Probably the cleanest signal with respect to chiral symmetry comes from the analysis of mesonic screening masses, which are relatively easy to compute in lattice simulations, see e.g.\ Ref.\ \cite{Cheng:2010fe}, even though their relation to phenomenologically relevant  quantities is not immediately clear (see however Ref.\ \cite{Brandt:2014uda}).
Direct computation of spectral quantities in a medium, such as thermal masses, is considerably harder, due to the need to consider analytical continuation on lattices with a finite temporal extent. Recent interesting work on the pion in the hadronic gas can be found in Refs.\ \cite{Brandt:2014qqa,Brandt:2015sxa}. 
The vector meson correlator has been analysed extensively, not only due its role in the dilepton rate 
but also in the context of the electrical conductivity and  the charge diffusion coefficient  \cite{Aarts:2007wj,Ding:2010ga,Amato:2013naa,Aarts:2014nba,Brandt:2015aqk,Ding:2016hua}.
Concerning quarkonia, both charmonium  \cite{Umeda:2002vr,Asakawa:2003re,Datta:2003ww,Aarts:2007pk,Ohno:2011zc,Borsanyi:2014vka} and, more recently, bottomonium \cite{Aarts:2011sm,Aarts:2013kaa,Aarts:2014cda,Kim:2014iga} have been studied on the lattice.

Surprisingly, even though light baryons are sensitive to chiral symmetry and play an important role in the analysis of heavy-ion data, corresponding studies in the baryonic sector are very limited. In the context of lattice QCD, baryon screening masses in a gluonic medium were studied a long time ago in Refs.\ \cite{DeTar:1987xb,DeTar:1987ar} and, at small baryon chemical potential, in Ref.\ \cite{Pushkina:2004wa}. More recently, screening and temporal correlators were analysed in Ref.\ \cite{Datta:2012fz}. All these studies were carried out in the quenched approximation.

In this work we aim to improve this situation substantially. We study the $N$ (nucleon), $\Delta$ and $\Omega$ baryons, employing simulations with $N_f=2+1$ light flavours, with four temperatures below and four above the transition. This allows us to study the properties of these baryons and in particular in-medium modification in the hadronic gas.
Chiral symmetry is closely linked to parity doubling and we analyse the emergence of parity doubling as the transition is approached. We find a qualitative difference in the response to the increasing temperature between positive- and negative-parity baryons, which may be of interest for heavy-ion phenomenology.
We also contrast the behaviour in the $\Omega$ channel with the $N$ and $\Delta$ channel, to see the effect of the heavier $s$ quark.

This paper is organised as follows. In Sec.\ \ref{sec:theory} we summarise the relations between baryon correlators and spectral functions, emphasising the differences with the mesonic case. We discuss positivity of the spectral functions, the role of charge conjugation, and the connection between chiral symmetry and parity doubling, both at $\mu=0$ and $\mu\neq 0$.
Sec.~\ref{sec:setup} contains details of our lattice computation. The main results of our study are given in Sec.\ \ref{sec:correlators}: we analyse the euclidean correlators and draw conclusions for both the hadronic gas and the quark-gluon plasma. These results are supported by the spectral function analysis of Sec.~\ref{sec:spec}. The final section summarises and contains an outlook.
We note here that our previous work in the nucleon sector, with limited statistics, can be found in Ref.~\cite{Aarts:2015mma} and preliminary results have appeared in Refs. \cite{Aarts:2015xua,Aarts:2016ioq,Aarts:2016kvt}.


\section{Baryonic correlators and spectral functions}
\label{sec:theory}

We start with a brief discussion of baryonic operators and spectral relations for fermionic  two-point functions. While for mesonic (bosonic) correlators the type of relations discussed below are very well known \cite{Bellac:2011kqa}, for fermionic ones this is slightly less so. Moreover, it allows us to discuss how parity doubling manifests itself in correlators and spectral functions.

\subsection{Baryonic operators}

We consider two-point functions of fermionic operators, of the form 
\be
G^{\alpha\alpha'}(x) = \avg{O^\alpha(x)\,\overline{O}^{\alpha'}(0)},
\ee
where $\alpha, \alpha'$ are Dirac indices and $\overline O = O^\dagger \gamma_4$.\footnote{We follow the conventions in Ref.\ \cite{Gattringer:2010zz} and use euclidean gamma-matrices, $\gamma_\mu^{\dagger}=\gamma_\mu=\gamma_\mu^{-1}$, with $\mu=1,\dots,4$, and $\gamma_5^\dagger=\gamma_5=\gamma_1\gamma_2\gamma_3\gamma_4$.}
The simplest annihilation operators for the nucleon, $\Delta$ and $\Omega$ baryons are respectively~\cite{Leinweber:2004it,Edwards:2004sx}
\be
\label{eq:ON}
O_N^{\alpha}(x) &=& \epsilon_{abc}\, u_a^{\alpha}(x) \left(d^{^T}_b(x) C \gamma_5 u_c(x) \right), \\
\label{eq:ODelta}
O_{\Delta,i}^{\alpha}(x) &=& 
 \epsilon_{abc} \left[  2 u_a^{\alpha}(x)\left(d^{^T}_b(x) C \gamma_i  u_c(x)\right) + d_a^{\alpha}(x)\left(u^{^T}_b(x) C \gamma_i u_c(x)\right) \right], \\
\label{eq:OOmega}
O_{\Omega,i}^{\alpha}(x) &=& \epsilon_{abc}\, s_a^{\alpha}(x)\left(s^{^T}_b(x) C \gamma_i s_c(x)\right),
\ee
where $C$ corresponds to the charge conjugation matrix, satisfying
\be
C^{\dagger}C = \Id, \qquad\quad \gamma_{\mu}^{^T} = -C \gamma_{\mu}C^{-1}, \quad\qquad C^{^T}=-C^{-1},
\ee
and hence $\gamma_5^{^T}  = C \gamma_5 C^{-1}$.
We note here that as written Eq.\ (\ref{eq:ODelta}) describes the charged $\Delta^+(uud)$ channel. However, since QED interactions are not incorporated and the two light quarks are taken to be degenerate (isospin limit), the operator is also relevant for  the neutral $\Delta^0(ddu)$ channel. The  $\Delta^{++}(uuu)$ and $\Delta^{-}(ddd)$ states are in principle described by an operator of the form (\ref{eq:OOmega}), with $s\to u,d$, but again in the degenerate limit one can show that Wick contractions coming from the latter are identical to the ones derived from Eq.\ (\ref{eq:ODelta}).

Under parity, elementary quark fields transform as
\be
\P \psi(x)\P^{-1} = \gamma_4\psi(Px),   \qquad\qquad P=\diag(-1,-1,-1,1).
\ee
It is straightforward to verify that this property is inherited by the baryonic operators,
\be
\P O(x)\P^{-1} = \gamma_4O(Px).
\ee
Hence one may introduce parity projectors and operators via
\be
\label{eq:Ppm}
 P_{\pm} = \half \left(\id \pm \gamma_4\right), 
 \qquad\quad
 O_\pm(x) = P_\pm O(x),
\ee
such that
\be
\P O_\pm(x)\P^{-1} = \pm O_\pm(Px).
\ee
We refer to $O_\pm$ as positive- and negative-parity operators. 

Similarly, under charge conjugation quark fields transform as
\be
\C \psi\C^{-1} \equiv \psi^{(c)} = C^{-1}\bar\psi^{^T}, 
 \qquad\qquad
 \C \bar\psi \C^{-1} \equiv \bar\psi^{(c)} = -\psi^{^T}C. 
\ee
Again, this is inherited by the baryonic operators, and
\be
\label{eq:OC}
 O^{(c)} = C^{-1}\overline O^{^T}, 
 \qquad\qquad
 \overline O^{(c)} = -O^{^T}C. 
\ee

\subsection{Spectral relations}

We now derive some general spectral relations and properties of the two-point functions $G^{\alpha\alpha'}(x)$.
  We work in spatial momentum space,
\be
G^{\alpha\alpha'}(\tau, \pv) = \int d^3x \, e^{-i\pv\cdot\xv} G^{\alpha\alpha'}(\tau,\xv),
\ee
and $\tau$ denotes the euclidean time, $0\leq \tau<1/T$, with $T$ the temperature. Fermionic fields and operators  satisfy anti-periodic boundary conditions in euclidean time.
A Fourier transform yields the correlator as a function of the fermionic Matsubara frequencies $\omega_n=(2n+1)\pi T$, $n\in\mathbb{Z}$, which can be written as a spectral integral
\be
\label{corr-spec}
G^{\alpha\alpha'} (i\omega_n,\pv) = \int_{-\infty}^{\infty} \frac{d\omega}{2\pi}\, \frac{\rho^{\alpha\alpha'}(\omega,\pv)}{\omega-i\omega_n}.
\ee
The spectral function $\rho^{\alpha\alpha'}(\omega,\pv)$ is then given by twice the imaginary part of the retarded Green function,
\be
\rho^{\alpha\alpha'}(\omega,\pv) = 2\Im G^{\alpha\alpha'}(i\om_n\to \om+i\eps, \pv),
\ee
or, in terms of the operators, by
\be
\rho^{\alpha\alpha'}(x) =  \avg{ \{ O^\alpha(x), \overline O^{\alpha'}(0) \} }, 
\ee
as always \cite{Bellac:2011kqa}. Transforming back to euclidean time yields the integral relation
\be
\label{eq:GK}
G^{\alpha\alpha'}(\tau, \pv) = \int_{-\infty}^{\infty} \frac{d\omega}{2\pi}\, K(\tau,\om) \rho^{\alpha\alpha'}(\omega,\pv),
\ee
with the kernel, for $0<\tau<1/T$,
\be
K(\tau,\om) = T\sum_n \frac{e^{-i\om_n\tau}}{\omega-i\omega_n} = \frac{e^{-\om\tau}}{1+e^{-\om/T}} = e^{-\om\tau}\left[ 1-n_F(\om)\right],
\ee
which can e.g.\ be shown by contour integration \cite{Bellac:2011kqa}.
Here $n_F(\om)=1/(e^{\om/T}+1)$ is the Fermi-Dirac distribution. We note that $K(\tau,\om)$ is neither even nor odd, but satisfies 
\be
\label{eq:Ktt}
 K(1/T-\tau,\om) = K(\tau,-\om).
 \ee
A decomposition of the kernel in terms of its even and odd parts yields
\be
 K(\tau,\om) = \half\left[ K_{\rm e}(\tau,\om) + K_{\rm o}(\tau,\om)\right],
 \ee
 with
\be
\label{eq:Keo}
\begin{aligned}
K_{\rm e}(\tau,\om) &=  \frac{\cosh(\om \tilde \tau)}{\cosh(\om/2T)} = \left[1-n_F(\om)\right] e^{-\om\tau} + n_F(\om)e^{\om\tau}, 
 \\
K_{\rm o}(\tau,\om) &=  -\frac{\sinh(\om \tilde \tau)}{\cosh(\om/2T)} = \left[1-n_F(\om)\right] e^{-\om\tau} - n_F(\om)e^{\om\tau},
\end{aligned}
\ee
where $\ttau=\tau-1/(2T)$. Note that the normalisation is such that all kernels reduce to $e^{-\om\tau}$ in the zero-temperature limit (for positive $\om$).
These kernels should be contrasted with the kernel appearing in bosonic spectral relations, 
\be
K_{\rm boson}(\tau,\om) &=&  \frac{\cosh(\om \tilde \tau)}{\sinh(\om/2T)} = \left[1+n_B(\om)\right] e^{-\om\tau} + n_B(\om)e^{\om\tau}, 
\ee
where  $n_B(\om)=1/(e^{\om/T}-1)$ is the Bose-Einstein distribution. The different denominators, $\cosh(\om/2T)$ versus $\sinh(\om/2T)$, reflect the quantum statistics. Note that as a consequence the problems associated with the singular behaviour  of the bosonic kernel, $K_{\rm boson}(\tau,\om)\to 2T/\om$ as $\om\to 0$, relevant for transport \cite{Aarts:2002cc}, are absent in the fermionic case.

In order to resolve the Dirac indices, we use the decomposition (other tensor structures will not appear in our application)
\bea
G^{\alpha\alpha'}(x) &=& \sum_\mu \gmu^{\alpha\alpha'} G_\mu(x) + \id^{\alpha\alpha'} G_m(x), \\
\rho^{\alpha\alpha'}(x) &=& \sum_\mu \gmu^{\alpha\alpha'} \rho_\mu(x) + \id^{\alpha\alpha'} \rho_m(x),
\eea
such that
\be
G_\mu(x) = \frac{1}{4}\tr \gmu G(x), \qquad\qquad G_m(x) = \frac{1}{4}\tr G(x),
\ee
where the trace is over the Dirac indices, and similarly for $\rho_{\mu,m}$.

Below we will specialise to zero spatial momentum, for which $G_i$ and $\rho_i$ vanish. It is convenient to combine the two remaining components with the help of positive- and negative-parity projectors (\ref{eq:Ppm}) as
\bea
G_\pm(x) &=& \tr P_\pm G(x) =  \tr\!\avg{O_\pm(x)\overline O_\pm(0)}  = 2\left[ G_m(x) \pm G_4(x) \right], \\
\rho_\pm(x) &=& \tr P_\pm \rho(x) =  \tr\!\avg{ \{ O_\pm(x), \overline O_\pm(0) \} }   = 2\left[ \rho_m(x) \pm \rho_4(x) \right],
\eea
related via
\be
\label{eq:Grho}
 G_\pm(\tau,\pv)  =  \int_{-\infty}^{\infty} \frac{d\omega}{2\pi}\, K(\tau,\om) \rho_\pm(\omega,\pv).
\ee
We will now prove a number of properties of $\rho_\pm(x)$ and $\rho_{4,m}(x)$.

\subsection{Positivity}
\label{sec:pos}

We start with positivity: we will show\footnote{Note that we use the notation $p=(p^0,\pv)$ with $p^0=\om$.} that $\pm\rho_\pm(p), \rho_{4}(p) \geq 0$ for {\em all} $\om$, while $\rho_m(p)$ does not have a definite sign, even when restricting to $\om\gtrless 0$.

Suppressing Dirac indices and using the KMS condition \cite{Bellac:2011kqa}, valid in thermal equilibrium,  we can write
\be
\label{eq:def}
\rho(p) &=& G^>(p)-G^<(p) \stackrel{\rm KMS}{=} \left( 1 + e^{-p^0/T} \right)G^>(p) \nn \\ 
&=& \left( 1 + e^{-p^0/T}\right)\int d^4x\, e^{-i p\cdot x}\, G^>(x),
\ee
 where  $G^{\lessgtr}$ are the usual Wightman functions \cite{Bellac:2011kqa},
 \be
 G^>(x-x') = \avg{ O(x) \overline O(x')}, 
 \qquad\qquad
  G^<(x-x') = -\avg{ \overline O(x') O(x) }.
\ee 
We first consider $\rho_4(p)$ and take the trace with $\gamma_4$. This yields
\be
\label{eq:rho4pos}
\rho_4(p) =  \left( 1 + e^{-p^0/T}\right)\int d^4x\, e^{-i p\cdot x}\,  \frac{1}{4}\tr \avg{ O(x) O^\dagger(0)}.
\ee
To proceed, we use the Heisenberg representation $O(x) =  e^{-ix\cdot K} O(0)  e^{ix\cdot K}$ and insert complete sets of eigenstates $|n\rangle$ of 
the translation operator $K^\mu$ (here $K^0$ is the Hamiltonian $H$ with eigenvalues $k^0_n$). Recalling that the expectation value denotes the thermal average with Boltzmann weight $e^{-H/T}/Z$, we find, after some rearrangement,
\be
  \rho_4(p) &=& \frac{1}{Z} \left( 1 + e^{-p^0/T} \right) 
   \sum_{n,m}  e^{- k^0_n/T} \,\frac{1}{4} \tr\! \left| \bra{n} O(0) \ket{m} \right|^2 \int d^4x\, e^{-i (p+k_n-k_m) \cdot x} \nn\\
   \label{eq:pos4}
&=& \frac{1}{Z} 
   \sum_{n,m,\alpha} \left( e^{- k^0_n/T} +  e^{- k^0_m/T}  \right) \frac{1}{4} \left| \bra{n} O^\alpha(0) \ket{m} \right|^2 (2\pi)^4 \delta^{(4)}(p+k_n-k_m), \;\;\;\;\;\;
\ee
 where we have written the Dirac index $\alpha$ explicitly again. It is easy to see that the terms added within the summation are nonnegative and hence we arrive at positivity: $\rho_4(p)\geq 0$ for all $p$.  

Next we consider $\rho_\pm(p)$ and take the trace with $P_\pm$. We now encounter
\be
\tr P_\pm O(x) \overline O(0) = \pm \tr O_\pm(x) O_\pm^\dagger(0),
\ee
where we used $\gamma_4=P_+-P_-, P_\pm^2=P_\pm, P_+P_-=0$, cyclicity of the trace, and Eq.~(\ref{eq:Ppm}). Proceeding as above then yields
\be
     \label{eq:pos}
     \rho_{\pm}(p) = \frac{\pm 1}{Z} 
   \sum_{n,m,\alpha} \left( e^{- k^0_n/T} +  e^{- k^0_m/T}  \right)  \left| \bra{n} O^\alpha_{\pm}(0) \ket{m} \right|^2 (2\pi)^4 \delta^{(4)}(p+k_n-k_m), \;\;\;\;\;\;
\ee
i.e.\ we find positivity of the spectral functions $\pm\rho_\pm(p)\geq 0$ for all $p$. 
 
 Positivity of $\rho_4(p)$ also follows from $\rho_4(p) = [\rho_+(p)-\rho_-(p)]/4\geq 0$; on the other hand,  $\rho_m(p) = [\rho_+(p)+\rho_-(p)]/4$ does not have a definite sign and is indeed not sign-definite for $\om\gtrless 0$, already at leading order in perturbation theory \cite{Praki:2015yua,Praki:inprep}.

To contrast, we note that for bosonic operators the spectral decomposition takes the form as above, but with a minus sign between the two thermal factors \cite{Bellac:2011kqa}. In addition, if the operator satisfies $J^\dagger = \pm J$,\footnote{This is e.g.\ the case for mesonic operators of the form $J=\bar\psi\Gamma\psi$, where $\Gamma$ is a Dirac matrix selecting the channel, since $J^\dagger = \bar\psi\gamma_4 \Gamma^\dagger \gamma_4\psi=\pm J$ \cite{Karsch:2003wy,Aarts:2005hg}.}
 it follows that the corresponding spectral function $\rho_B(p)$ is odd under $\om\to -\om$, 
 and $\om\rho_B(\om,\pv)\geq 0$. This can be seen by swapping $n\leftrightarrow m$ in the summation. For the fermionic operators we consider here, this argument does not apply, since $O^\dagger\neq \pm O$.
Hence in general fermionic spectral functions are neither even nor odd.

\subsection{Charge conjugation}

Next we relate, in the case of vanishing baryon chemical potential (or baryon density), positive- and negative-parity correlators and spectral functions, i.e.\ we show that 
\be
\label{eq:Gpmmp}
G_\pm(\tau,\pv) = -G_\mp(1/T-\tau,\pv), \qquad\qquad
\rho_{\pm}(-\om,\pv) =  -\rho_{\mp}(\om,\pv).
\ee

We follow Ref.\ \cite{Karsch:2009tp}, where this is demonstrated at the level of the single-quark propagator.
Here we consider baryonic (or fermionic in general) operators,  transforming under charge conjugation as in Eq.\ (\ref{eq:OC}). We assume isotropy, i.e.\ invariance under $\pv\to -\pv$, throughout.

The time-ordered correlation function is given by
\be
G^{\alpha\alpha'}(x-x') = \avg{ \mbox{T}_\tau \left[ O^\alpha(x) \overline O^{\alpha'}(x') \right] },
\ee
with the imaginary-time-ordered product  
\be
\mbox{T}_\tau\left[ A(\tau)B(\tau')\right] \equiv \theta(\tau-\tau')A(\tau)B(\tau') \pm \theta(\tau'-\tau)B(\tau')A(\tau).
\ee
Here the minus (plus) sign applies to fermionic (bosonic) operators. 

At zero chemical potential, thermal expectation values are invariant under charge conjugation. We hence find, suppressing Dirac indices,
\be
G(x-x') &=& \avg{ \mbox{T}_\tau \left[ \C O(x) \overline O(x')  \C^{-1}\right] } = 
 \avg{ \mbox{T}_\tau \left[ O^{(c)}(x) \overline O^{(c)}(x') \right] }
\nn \\
&=& - \avg{ \mbox{T}_\tau \left[ C^{-1} \overline O^{^T}(x) O^{^T}(x') C \right] }
=   \avg{  C^{-1} \mbox{T}_\tau \left[    \left( O(x')  \overline O(x) \right)^{^T}  \right]  C }
\nn \\
\label{eq:GT}
&=& C^{-1} G^{^T}\!(x'-x) C.
\ee
From now on we take $x=(\tau,\xv)$,  $x'=(0,\xv')$, with $0<\tau<1/T$. Using the cyclicity of thermal expectation values \cite{Bellac:2011kqa} then gives 
\be
G(x'-x) = G(-\tau, \xv'-\xv) = -G(1/T-\tau, \xv'-\xv).
\ee
Applying this to Eq.\ (\ref{eq:GT}), we find, in momentum space, 
\be 
G(\tau, \pv) = -C^{-1} G^{^T}\!(1/T-\tau, \pv) C.
\ee
We can now take the trace with $P_\pm$, which yields
\be 
G_\pm(\tau, \pv) &=&   \tr P_\pm G(\tau, \pv) 
 =  -\tr   P_\pm  C^{-1} G^{^T}\!(1/T-\tau, \pv) C 
 \nn\\
 &=&   -\tr  (C P_\pm C^{-1})^{^T}  G(1/T-\tau, \pv) =  -\tr  P_\mp  G(1/T-\tau, \pv) \nn\\
 &=&  -G_\mp(1/T-\tau, \pv) 
 \ee
where we used that 
\be
  (C P_\pm C^{-1})^{^T}  = P_\mp.
\ee
We have now demonstrated the first relation in Eq.\ (\ref{eq:Gpmmp}). The second relation immediately follows, when using the integral relations (\ref{eq:Grho}) as well as the property (\ref{eq:Ktt}).
Physically it reflects that positive-parity states propagate forward in euclidean time, when using $G_+$, and backward in time when using  $G_-$, and vice versa for negative-parity states. In terms of spectral functions, this relates the positive part of the spectrum of $\rho_+$ with the negative part of $\rho_-$, and again vice versa.  Explicitly, if the spectrum is dominated by single groundstates with masses $m_\pm$, this implies
\be
\label{eq:Gsimple}
\pm G_\pm(\tau)= A_{\pm} e^{-m_\pm\tau} + A_\mp e^{-m_\mp(1/T-\tau)}.
\ee

Using the relation $\rho_{\pm}(-p) =  -\rho_{\mp}(p)$, we can subsequently note that 
\be
\rho_4(p)  &=& \frac{1}{4} \left[\rho_+(p) - \rho_-(p) \right] = \frac{1}{4} \left[\rho_+(p) + \rho_+(-p) \right], \\
\rho_m(p) &=& \frac{1}{4} \left[\rho_+(p) + \rho_-(p) \right] = \frac{1}{4} \left[\rho_+(p) -\rho_+(-p) \right], 
\ee 
and hence these are even, respectively odd under $\om\to -\om$.
Their spectral relations hence involve the even and odd kernels $K_{\rm e,o}(\tau,\om)$ respectively, see Eq.\ ({\ref{eq:Keo}}).
We remark that this only holds when there is no net density, i.e.\ when the density matrix is invariant under charge  conjugation.

\subsection{Chiral symmetry and parity doubling}

The final relations we derive are for the case of unbroken chiral symmetry. Here we work in the harmonious  world of thermal field theory in which chiral symmetry is simply expressed as $\{\gamma_5, G\}=0$, sidestepping momentarily the issues related to chiral symmetry in realistic lattice QCD computations, to be discussed below.

From the anti-commutation relation of the correlator with $\gamma_5$, it immediately follows that
\be
G_m(x)=\rho_m(x)=0,
\ee
and hence
\be
G_+(\tau,\pv) = - G_-(\tau,\pv)  &=& G_+(1/T-\tau,\pv) = 2G_4(\tau,\pv), \\
\label{eq:rhopd}
\rho_+(p) = -\rho_-(p) &=&  \rho_+(-p) = 2\rho_4(p).
\ee
These relations imply that the lattice correlators are symmetric around the centre of lattice ($\tau=1/2T$), the spectral functions are even functions in $\om$ and that identical spectral information is contained in $\rho_\pm(p)$. We refer to this as parity doubling.
We emphasise that any of these signatures  are equivalent statements of parity doubling.

An alternative proof for two massless flavours goes as follows \cite{Gattringer:2010zz}. When chiral symmetry is unbroken, i.e.\ the quarks are massless and chiral symmetry is not broken spontaneously, the theory is unchanged when the following chiral rotation is performed on the quark fields, 
\be
\psi \to \exp(i\alpha\gamma_5 T_3)\psi, \qquad\qquad  \bar\psi \to \bar\psi \exp(i\alpha\gamma_5 T_3),
\ee
where $\psi=(u,d)^T$ and $T_3=\sigma_3/2$ acts in flavour-space. Choosing $\alpha=\pi$, we get distinct chiral transformations on the spinor fields:
\be
\label{eq:chiral}
u \to i\gamma_5 u,  \qquad \bar u \to i\bar u \gamma_5, \qquad d \to -i\gamma_5 d, \qquad \bar d \to -i\bar d\gamma_5.
\ee
It is then easy to check that the $N$ and $\Delta$ operators, see Eqs.\ (\ref{eq:ON}, \ref{eq:ODelta}), transform as
\be
 O_N \to i\gamma_5 O_N,
\qquad\qquad
  O_\Delta \to -i \gamma_5 O_\Delta,
 \ee
and hence
\be
O_\pm^N =  P_\pm O_N  \to  i\gamma_5 O^N_\mp,
\qquad\quad
O_\pm^\Delta =  P_\pm O_\Delta \to -i \gamma_5 O^\Delta_\mp.
\ee
In both channels the correlator then transforms as
\be
G_{\pm}(x) = \tr\! \avg{ O_\pm(x) \overline O_\pm(0)} 
\to 
- \tr\! \avg{ O_\mp(x) \overline O_\mp(0)} = -G_\mp(x),
\ee
which was to be shown.

\subsection{Nonzero chemical potential}

For completeness, we indicate here how the properties derived above are modified in presence of a nonzero baryon chemical potential $\mu$, such that the Hamiltonian in the Boltzmann weight is changed from $H\to H-\mu Q$, with $Q$ the baryon number.
 
 First we consider positivity. Following the same steps as in Sec.\ \ref{sec:pos}, in which the KMS condition (\ref{eq:pos4}) is modified as $p^0\to p^0-\mu$, and using that the states $\ket{n}$ are simultaneous eigenstates of  $H$ and $Q$ (with eigenvalues $q_n$), we arrive at
\be
  \rho_4(p) &=& \frac{1}{Z} 
   \sum_{n,m} \left( e^{- (k^0_n-\mu q_n)/T} +  e^{- (k^0_m-\mu q_m)/T}  \right)
  \nn\\ && 
    \frac{1}{4}\tr\! \left| \bra{n} O(0) \ket{m} \right|^2 (2\pi)^4 \delta^{(4)}(p+k_n-k_m), 
\ee
and similar for $\rho_\pm(p)$. 
Hence positivity holds, as before.

At nonzero chemical potential, the density matrix is not invariant under charge conjugation, since baryon number changes sign. Therefore invariance is obtained by simultaneously changing $\mu\to -\mu$, which yields the relations
\be
G_\pm(\tau,\pv;\mu) &=& -G_\mp(1/T-\tau,\pv;-\mu), \\
\rho_{\pm}(-\om,\pv;\mu) &=&  -\rho_{\mp}(\om,\pv;-\mu).
\ee
$G_{4,m}$ are then no longer (anti)symmetric around $\tau=1/2T$, but satisfy instead
\be
G_4(1/T-\tau,\pv;\mu) &=& G_4(\tau,\pv; -\mu), \\
G_m(1/T-\tau,\pv;\mu) &=& -G_m(\tau,\pv; -\mu).
\ee
 Again explicitly, if the spectrum is dominated by single groundstates, Eq.\ (\ref{eq:Gsimple}) is modified as
\be
G_+(\tau;\mu) &=& A_+(\mu) e^{-(m_+-\mu)\tau} + A_-(-\mu) e^{-(m_-+\mu)(1/T-\tau)},
\\
-G_-(\tau;\mu) &=& A_-(\mu) e^{-(m_--\mu)\tau} + A_+(-\mu) e^{-(m_++\mu)(1/T-\tau)}.
\ee
Finally, in the case of unbroken chiral symmetry, $G_m(x)=\rho_m(x)=0$ still holds and 
\be
G_+(\tau,\pv;\mu) = - G_-(\tau,\pv;\mu)  &=& G_+(1/T-\tau,\pv;-\mu)   = 2G_4(\tau,\pv;\mu), \\
\rho_+(p;\mu) = -\rho_-(p;\mu) &=&   \rho_+(-p;-\mu) = 2\rho_4(p;\mu).
\ee


\section{Lattice setup}
\label{sec:setup}

\begin{table}[t]
\begin{center}
\begin{tabular}{cccccccc}
\hline
$N_s$ & $N_\tau$  & $T$\,[MeV] & $T/T_c$  & $N_{\rm src}$  & $N_{\rm cfg}$\\
\hline
  24 & 128 & 44	& 0.24  &  16& 139\\
  24 & 40 & 141 & 0.76  &  4  & 501  \\
  24 & 36 & 156 & 0.84  &  4  & 501  \\
  24 & 32 & 176 & 0.95  &  2  & 1000 \\ 
  24 & 28 & 201 & 1.09  &  2  & 1001 \\ 
  24 & 24 & 235 & 1.27  &  2  & 1001 \\  
  24 & 20 & 281 & 1.52  &  2  & 1000 \\ 
  24 & 16 & 352 & 1.90  &  2  & 1001 \\ 
\hline
\end{tabular}
\caption{Ensembles used in this work. The lattice size is $N_s^3 \times N_\tau$ , with the temperature $T = 1/(a_\tau N_\tau)$.
The available statistics for each ensemble is $N_{\rm cfg} \times N_{\rm src}$. The sources were chosen randomly in the four-dimensional lattice. The spatial lattice spacing $a_s = 0.1227(8)$ fm, the inverse temporal lattice spacing $a^{-1}_\tau = 5.63(4)$ GeV, and the renormalised anisotropy $\xi=a_s/a_\tau =3.5$.
}
\label{tab:lat}
\end{center}
\end{table}

We have computed baryon correlators using the thermal ensembles of the FASTSUM collaboration \cite{Amato:2013naa,Aarts:2014nba,Aarts:2014cda}. These ensembles are generated with $2+1$ flavours of Wilson fermions on an anisotropic lattice, with a smaller temporal lattice spacing, $a_\tau<a_s$; the renormalised anisotropy is $\xi\equiv a_s/a_\tau =3.5$. 
The lattice action used is the Symanzik-improved anisotropic gauge action with tree-level mean-field coefficients and a mean-field-improved Wilson-clover fermion action with stout-smeared links and follows the Hadron Spectrum Collaboration \cite{Edwards:2008ja}. Details of the action and parameter values can be found in Refs.\ \cite{Aarts:2014cda,Aarts:2014nba}.
The choice of masses for the degenerate $u$ and $d$ quarks yields a pion with a mass of $M_\pi=384(4)$ MeV \cite{Lin:2008pr}, which is heavier than in nature, while the strange quark has been tuned to its physical value.
Configurations and correlation functions have been generated using the CHROMA software package \cite{Edwards:2004sx}, via the SSE
optimizations when possible \cite{McClendon}.

We use a fixed-scale approach, in which the temperature is varied by changing $N_\tau$, according to $T = 1/(a_\tau N_\tau)$. 
Table \ref{tab:lat} gives an overview of the ensembles. Access to the ``zero-temperature'' configurations ($N_\tau = 128$) has been kindly provided to us by the Hadron Spectrum Collaboration. 
An estimate for the pseudo-critical temperature, $T_c=185(4)$ MeV,  follows from an analysis of the renormalized Polyakov loop  \cite{Aarts:2014nba} and is higher than in nature, due to the large pion mass.
Note that there are four ensembles in the hadronic phase and four in the quark-gluon plasma.

Concerning the baryonic correlators, Gaussian smearing \cite{Gusken:1989ad} has been employed to increase the overlap
with the groundstate. In order to have a positive spectral weight, we apply the
smearing on both source and sink, i.e., 
\be
\psi' = \frac{1}{A} \left(\id + \kappa H \right)^{n} \psi, 
\ee
where $A$ is a normalisation factor and $H$ is the spatial hopping part of the Dirac operator.
The hopping term contains APE smeared links \cite{Albanese:1987ds} using $\alpha =1.33$ and one iteration. 
We tuned the parameters to the values $n=60$ and $\kappa = 4.2$, by maximising
the length of the plateau for the effective mass of the groundstate at the lowest temperature. 
Smearing is applied only in the spatial directions, equally to all temperatures and ensembles.


\section{Thermal baryon correlators}
\label{sec:correlators}

In this section we present the results for the baryon correlators at all temperatures. Based on the determination of the pseudo-critical temperature $T_c$ via the renormalised Polyakov loop, the discussion is organised in terms of the hadronic gas ($T<T_c$) and the quark-gluon plasma ($T>T_c$). Since the transition is a crossover, it is not immediately obvious at which temperatures light and strange baryons  cease to exist.\footnote{We remind the reader that the temperatures closest to $T_c$ are $T/T_c=0.95$ and $1.09$.} However, below we will find clear indications that the baryonic bound states are absent at $T/T_c=1.09$, in the three  channels we consider.

\subsection{Hadronic gas}

\begin{figure}[t]
\centering
\includegraphics[width=0.7\textwidth]{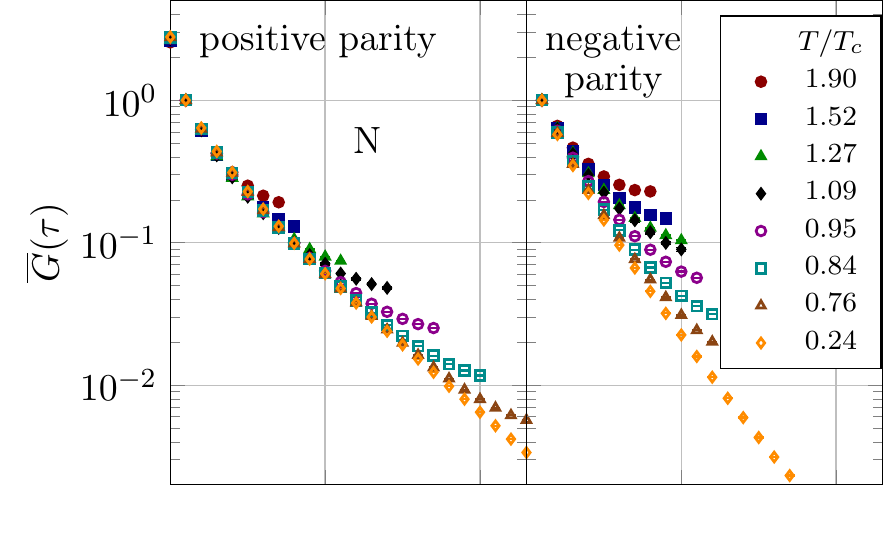}

\vspace*{-0.65cm}

\includegraphics[width=0.7\textwidth]{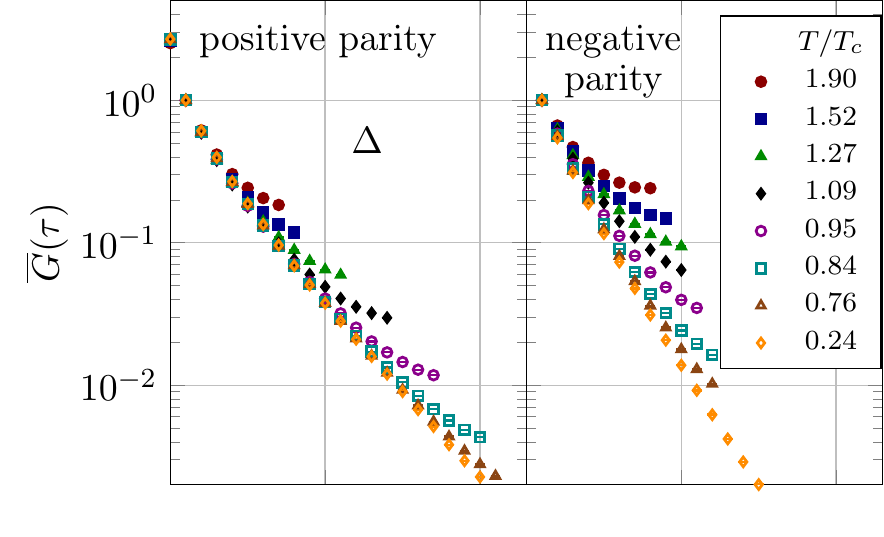}

\vspace*{-0.65cm}

\includegraphics[width=0.7\textwidth]{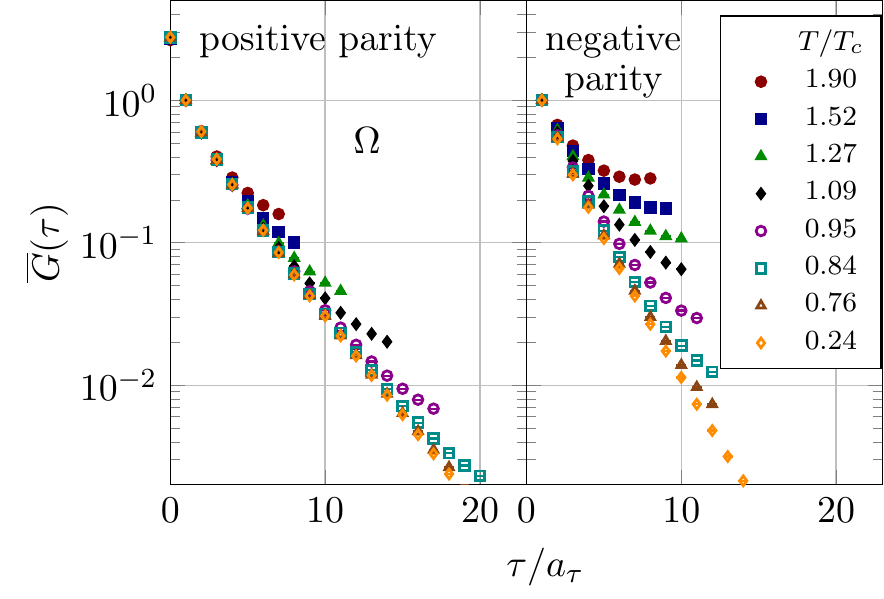}
\caption{Correlators $\overline G(\tau)$ for the positive- and negative-parity channels at different temperatures, in the $N$,  $\Delta$ and $\Omega$ sectors, on a logarithmic scale
}
\label{fg:nuclcorr}
\end{figure}

We have computed the baryon two-point functions in the $N$, $\Delta$ and $\Omega$ channels on the lattice ensembles  discussed above, at zero spatial momentum $\pv=\vecnul$ (we drop the momentum labels from now on). The results are shown in Fig.~\ref{fg:nuclcorr}, at all the eight temperatures available. 
The positive- and negative-parity channels are shown separately, i.e.\ the negative-parity channel is obtained using Eq.\ (\ref{eq:Gpmmp}),
and the correlators are normalised to the first Euclidean time point,  $\tau/a_\tau=1, N_\tau-1$ respectively, such that
\be
 \overline{G}_{+}(\tau) = \frac{G_+(\tau)}{\langle G_+(a_\tau) \rangle},
 \qquad\qquad
\overline{G}_{-}(\tau) = \frac{G_+( N_\tau a_\tau- \tau)}{\langle G_+(N_\tau a_\tau - a_\tau) \rangle}.
\ee

At low temperatures (open symbols), the correlators show exponential decay, indicating the presence of a well-defined groundstate. As the temperature is increased, some temperature dependence on the positive-parity side is observed, but considerably more temperature dependence is visible on the negative-parity side. The correlators naturally bend upwards around the minima, which are, however, not in the centre of the lattice ($\tau/a_\tau=N_\tau/2$), due to the absence of parity doubling, see e.g.\ Eq.\ (\ref{eq:Gsimple}).
Above $T_c$ (filled symbols), the correlators appear to drop slower than  exponential, indicating the absence of a well-separated groundstate. 

To analyse this quantitatively, we have fitted the correlators to a combination of simple exponentials, see Eq.\ (\ref{eq:Gsimple}), 
\be
G_+(\tau)= A_{+} e^{-m_+\tau} + A_- e^{-m_-(1/T-\tau)},
\label{eq:ansatz}
\ee
with $m_\pm$ the groundstate masses in both parity channels.  
While Fig.\ \ref{fg:nuclcorr} shows the positive- and negative-parity channels separately, the fit is carried out to the correlator $G_+(\tau)$ in one go. Around the minimum of the correlator, one might become susceptible to signal-to-noise problems, but we found this to be relevant at the lowest temperature only. Here we excluded points around the minimum of the correlators from the analysis, based on the quality of the fit and error analysis.\footnote{For example, in the nucleon channel we included the intervals $\tau/a_\tau=5-40, 105-124$.}
In order to estimate the systematic uncertainties of the four fit parameters, we have considered various Euclidean time intervals and, to suppress contributions from excited states, we have excluded very small times.
We used the so-called Extended Frequentist Method~\cite{Yao:2006px,Durr:2008zz} to  carry out the statistical analysis: 
this method considers all possible variations and weighs the final results according to the obtained $p$-value, which measures how extreme an outcome is, see Refs.\ \cite{Yao:2006px,Durr:2008zz}  for more details.
In the confined phase we found that it is possible to extract the mass parameters $m_\pm$, whereas above $T_c$ the exponential fits are no longer adequate, as can be expected in the deconfined phase (see below).

\begin{table}[t]
\centering
\begin{tabular}{ccccc | c}
$T/T_c$ & 0.24 & 0.76 & 0.84 & 0.95 & PDG ($T=0$) \\
\midrule
$m^N_+$ [MeV]  & 1159(13) & 1192(39) & 1169(53) & 1104(40) & 939 \\
$m^N_-$ [MeV]  & 1778(52) & 1628(104) & 1425(94) & 1348(83) & 1535(10) \\
\midrule
$m^\Delta_+$ [MeV]  & 1459(58) & 1521(43) & 1449(42) & 1377(37) & 1232(2) \\
$m^\Delta_-$ [MeV]  & 2138(117) & 1898(106) & 1734(97) & 1526(74) & 1710(40)\\
\midrule
$m^\Omega_+$ [MeV]  & 1661(21) & 1723(32) & 1685(37) & 1606(43) & 1672.4(0.3) \\
$m^\Omega_-$ [MeV]  & 2193(30) & 2092(91) & 1863(76) & 1576(66) & 2250--2380--2470 \\
\midrule
$\delta m_N$ [MeV]       & 619(54) & 436(111) & 256(108) & 244(92) & 596(10) \\
$\delta m_\Delta$ [MeV]  & 679(131) & 377(114) & 285(106) & 149(83) & 478(40) \\
$\delta m_\Omega$ [MeV]  & 532(37) & 369(96) & 178(85) & -30(79) & 578--708--798\\
\midrule
$\delta_N$  & 0.211(19) & 0.155(35) & 0.099(40) & 0.100(35) & 0.241(1) \\
$\delta_\Delta$ & 0.189(37) & 0.110(31) & 0.089(31) & 0.051(28) & 0.162(14)  \\
$\delta_\Omega$  & 0.138(10) & 0.097(23) & 0.050(23) & -0.009(25) & 0.147--0.175--0.192 \\
\midrule
\end{tabular}
\caption{Groundstate masses $m_\pm$ in both parity sectors in the $N$, $\Delta$ and $\Omega$ channels below $T_c$. Estimates for statistical and systematic uncertainties are included. The final column shows the $T=0$ values in nature \cite{Agashe:2014kda}. Note that there are several candidates for $m^\Omega_-$.
The difference $\delta m_{N, \Delta, \Omega}$ is defined as $\delta m=m_--m_+$ and the dimensionless ratio $\delta_{N, \Delta, \Omega}$ as $\delta=(m_--m_+)/(m_-+m_+)$.
}
\label{tab:mass}
\end{table}

 Table \ref{tab:mass} lists the results for the masses $m_\pm$ in all three channels, at the four temperatures below $T_c$. The results are shown in units of MeV, using the estimate for the temporal lattice of $a_\tau^{-1}=5.63(4)$ GeV \cite{Edwards:2008ja}.
 Also shown are the PDG \cite{Agashe:2014kda} values at $T=0$. Since our light quarks are heavier than in nature, the groundstate masses in the $N$ and $\Delta$ channels at the lowest temperature are larger as well. The splitting between the positive- and negative-parity groundstate masses, denoted with $\delta m$, is of the right order, however. The strange quark mass is tuned to the physical value \cite{Lin:2008pr} and the result for the $\Omega_+$ mass is consistent with the PDG value (within errors). Surprisingly, the $\Omega_-$ particle has not been unambiguously  identified in the PDG and there are three candidates.
The value we obtain at $T=0.24T_c$ seems to favour the candidate with the lowest mass, but 
a systematic analysis (continuum extrapolation and physical $u$ and $d$ quarks) is necessary to make a more stringent prediction.
Our results for the spectrum at the lowest temperature are in agreement with those of the HadSpec collaboration for the positive-parity states \cite{Lin:2008pr}; for the negative-parity baryons the masses obtained in Ref.\ \cite{Edwards:2012fx} on a smaller spatial lattice ($16^3$ instead of $24^3$) are somewhat lower, at the $2\sigma$ level.

\begin{figure}[t]
\centering
\includegraphics[width=0.48\textwidth]{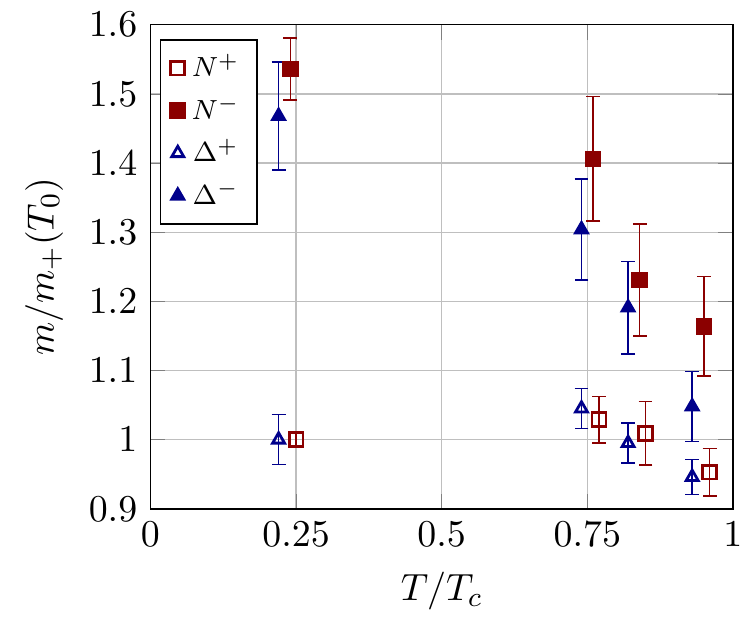}
\includegraphics[width=0.48\textwidth]{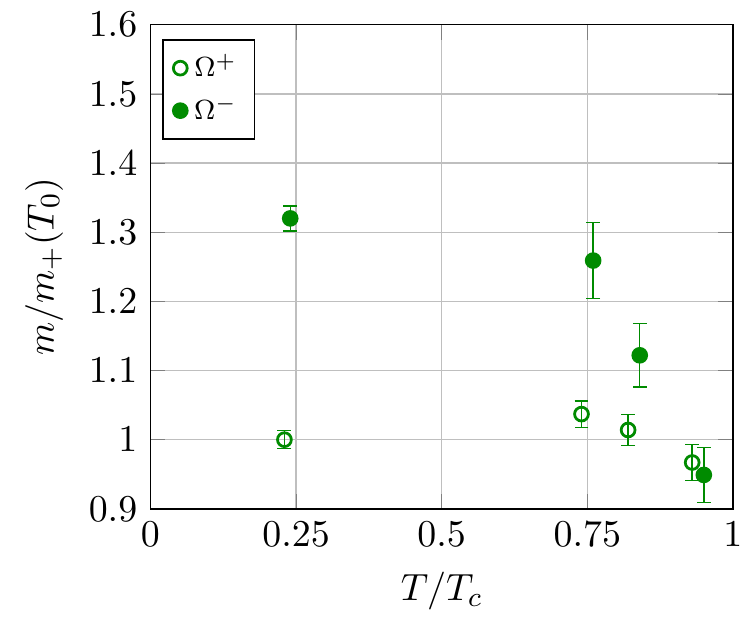}
\caption{Temperature dependence of $m_\pm$ of the $N$ and $\Delta$ (left, slightly shifted horizontally for clarity) and $\Omega$ (right) baryon, below $T_c$. The masses are normalised by $m^+$ at the lowest temperature, $T_0=0.24T_c$, in the channel under consideration.
}
\label{fig:masses}
\end{figure}

As the temperature is increased, we find that the groundstate mass in the positive-parity channels is largely unaffected by temperature; the deviation from the results at the lowest temperature is always less than 5\%.  Very close to $T_c$, the values drops slightly below the ones at $T/T_c=0.24$. This is further illustrated in Fig.\ \ref{fig:masses}, where the data are plotted normalised by $m_+$ at the lowest temperature, in the channel under consideration.
In the negative-parity channel we observe a stronger temperature dependence, which is remarkably similar in all three channels. Already at $0.75T_c$, the masses have dropped noticeably (see again Fig.\ \ref{fig:masses}) and this trend continues towards $T_c$. Very close to $T_c$ the parity channels are nearly degenerate. This is further quantified by the dimensionless ratio 
\be
\delta \equiv \frac{m_- - m_+}{m_- + m_+},
\label{eq:diff}
\ee
also included in Table  \ref{tab:mass}. The smaller value of $\delta_\Omega$ at all four temperatures is due to both $m_+^\Omega$ being larger and $\delta m_\Omega$ being smaller. Both of these effects are presumably due to the $s$ quark being heavier than the $u$ and $d$ quarks, which makes the contribution  to the groundstate mass due to chiral symmetry breaking less important in the $\Omega$ channel.

\subsection{Quark-gluon plasma}

\begin{figure}[t]
\centering
\includegraphics[width=0.48\textwidth]{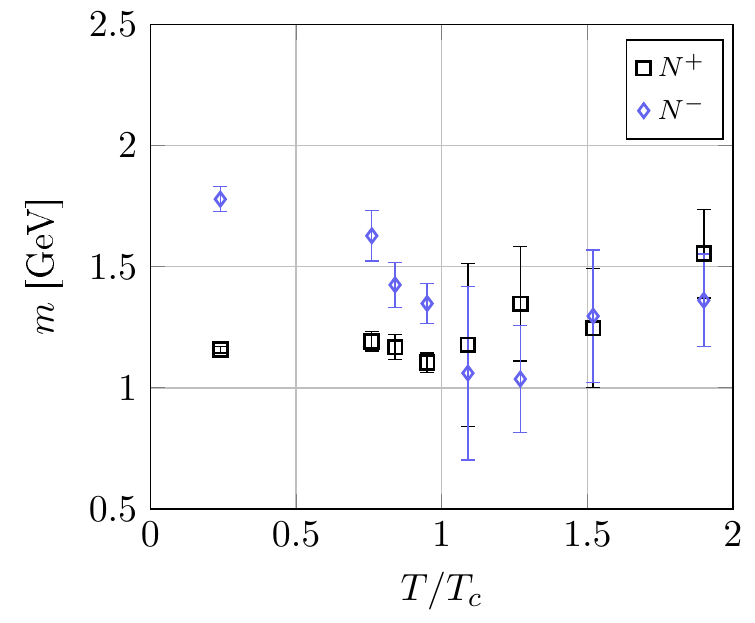} 
\includegraphics[width=0.48\textwidth]{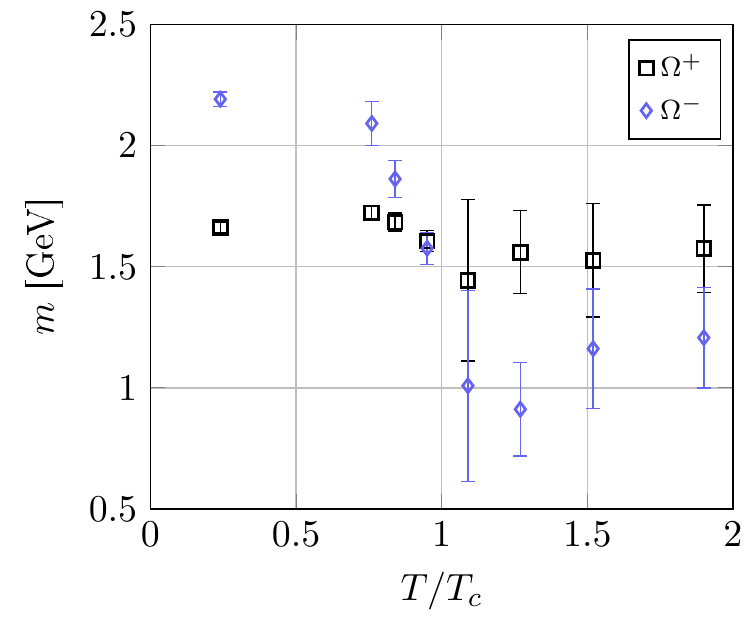}
\caption{Groundstate masses in the positive- and negative-parity channels at {\em all} temperatures, assuming the exponential decay of Eq.\ (4.2), in the $N$ (left) and $\Omega$ (right) channels.
}
\label{fig:masses-all}
\end{figure}

We now turn to the temperatures above the deconfinement transition. To start, we have considered the same analysis as above, using exponential fits, assuming that the hypothesis of separated well-defined groundstates still holds. The results are shown in Fig.\ \ref{fig:masses-all}, in the $N$ and $\Omega$ channels. We observe a clear qualitative change when going from $T/T_c=0.95$ to 1.09 (or reducing $N_\tau$ from 32 to 28). The error on the {\em would-be} groundstate masses, obtained by combining systematic and statistical uncertainties, is substantially larger, which cannot be simply explained by the reduction in the number of time slices used in the fits. This, and other results presented below, lead us to conclude that bound states are absent at $T/T_c=1.09$, both for the light baryons and in the $\Omega$ channel. Hence even though the transition is a crossover, we find that the spectrum changes rather drastically between 0.95 and 1.09$T_c$.

\begin{figure}[t]
\centering
\includegraphics[width=0.48\textwidth]{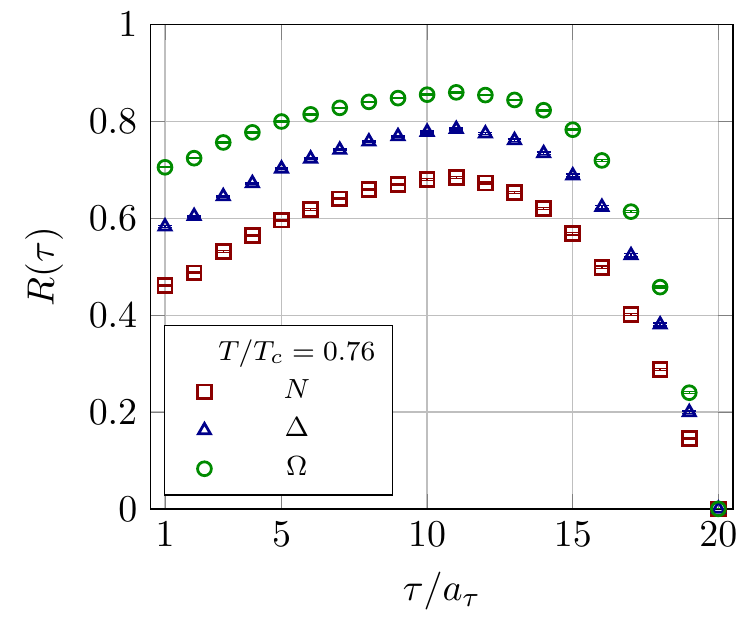}
\includegraphics[width=0.48\textwidth]{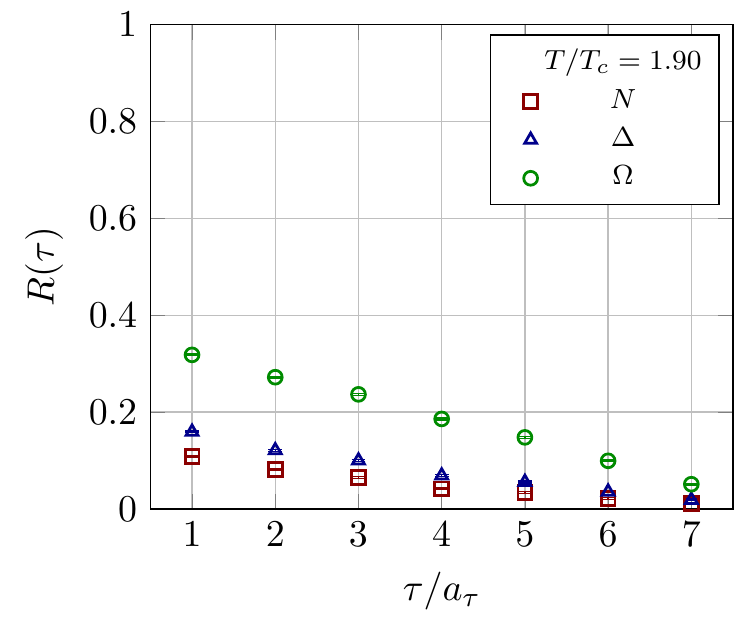}
\caption{Ratio $R(\tau)$ in the $N$, $\Delta$ and $\Omega$ channels, at $T/T_c=0.76$ (left) and 1.90 (right).
}
\label{fig:Rtau}       
\end{figure}

We hence focus on the signal for parity doubling, i.e.\ the emergent degeneracy in the positive- and negative parity channels. Following Ref.\ \cite{Datta:2012fz}, we study the ratio 
\be
R(\tau) =  \frac{G_+(\tau)-G_+(1/T-\tau)}{G_+(\tau)+G_+(1/T-\tau)},
\ee
which approaches 1 in the case that separated groundstates dominate, with $m_- \gg m_+$, but vanishes in the case of parity doubling. We have previously shown $R(\tau)$ for all temperatures in the nucleon sector \cite{Aarts:2016ioq}.
Here we present the outcome at two selected temperatures in Fig.\ \ref{fig:Rtau} in the $N$, $\Delta$ and $\Omega$ channels. We note the clear qualitative and quantitative difference: below $T_c$ the ratio is significantly different from zero,\footnote{We note that by construction $R(\tau)$ approaches zero at the centre of the lattice, $\tau/a_\tau=N_\tau/2$.}
while at the highest temperature it is much smaller. 
It should be emphasised that if chiral symmetry is exactly restored, complete degeneracy in the positive- and negative-parity channels is expected and $R(\tau)=0$. In our lattice simulations such a clear signal cannot be expected for a number of reasons. First of all we use Wilson fermions, which break chiral symmetry at short distances. We have found that smearing suppresses these contributions, yielding a better signal for parity doubling  \cite{Aarts:2016ioq}. Moreover, the quarks are not massless, with the two light flavours heavier than in nature. Hence this explicit symmetry breaking also affects the signal. However, this is expected to become less important at higher temperature, being suppressed as $m_q/T$. 
The effect of the finite quark mass can be seen in the splitting of $R(\tau)$ in Fig.~\ref{fig:Rtau} (right) between the $N, \Delta$ channels and the $\Omega$ channel at the highest temperature; this is most likely due to the larger $s$ quark mass.

\begin{figure}[t]
\centering
\includegraphics[width=0.6\textwidth]{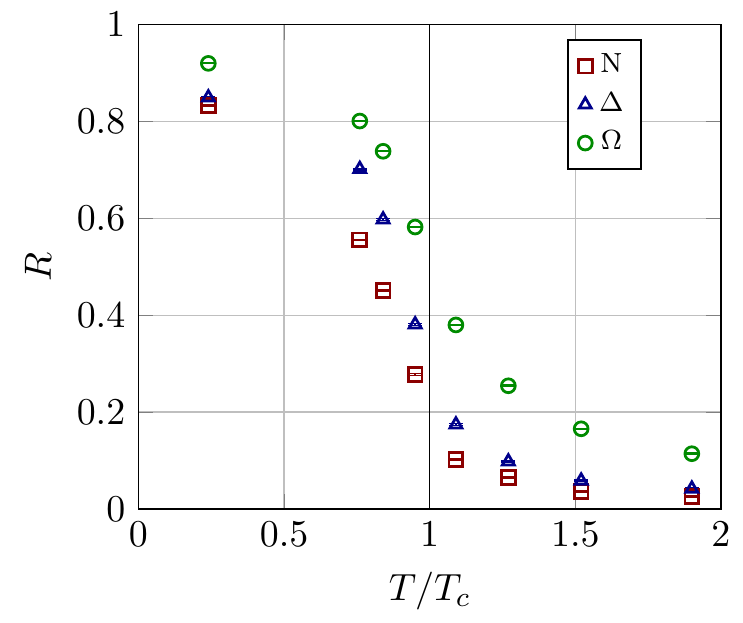}
\caption{Temperature dependence of $R$ in the $N$, $\Delta$ and $\Omega$ channels.}
\label{fig:R} 
\end{figure}

In order to summarise the results for all temperatures, we show in Fig.\ \ref{fig:R} the summed ratio
\be
\label{eq:ratioR}
R \equiv \frac{\sum_{n=1}^{N_\tau/2-1} R(\tau_n)/\sigma^2(\tau_n)}{\sum_{n=1}^{N_\tau/2-1}1/\sigma^2(\tau_n)},
\ee
where $\tau_n=a_\tau n$ and $\sigma(\tau_n)$ are the statistical uncertainties, used as weights.

Assuming that $m_+<m_-$, $R$ lies between 0 and 1, with $R=0$ corresponding to a symmetric correlator and parity doubling. We observe clear crossover behaviour in all three channels. The location of this transition is consistent with $T_c$, which has been determined by an analysis of the renormalised Polyakov loop. Hence it is natural to associate the transition with the approximate restoration of chiral symmetry in the quark-gluon plasma and to interpret $R$ as a quasi-order parameter. We also note that the effect is less pronounced in the $\Omega$ channel, due to the larger $s$ quark mass. It will therefore be interesting to study the effect of strangeness on parity doubling.
At the highest temperature available, $R>0$ in the $\Omega$ channel; it is expected that the effect of the quark mass will eventually disappear as $m_s/T\to 0$.

\begin{figure}[t]
\centering
\includegraphics[width=0.68\textwidth]{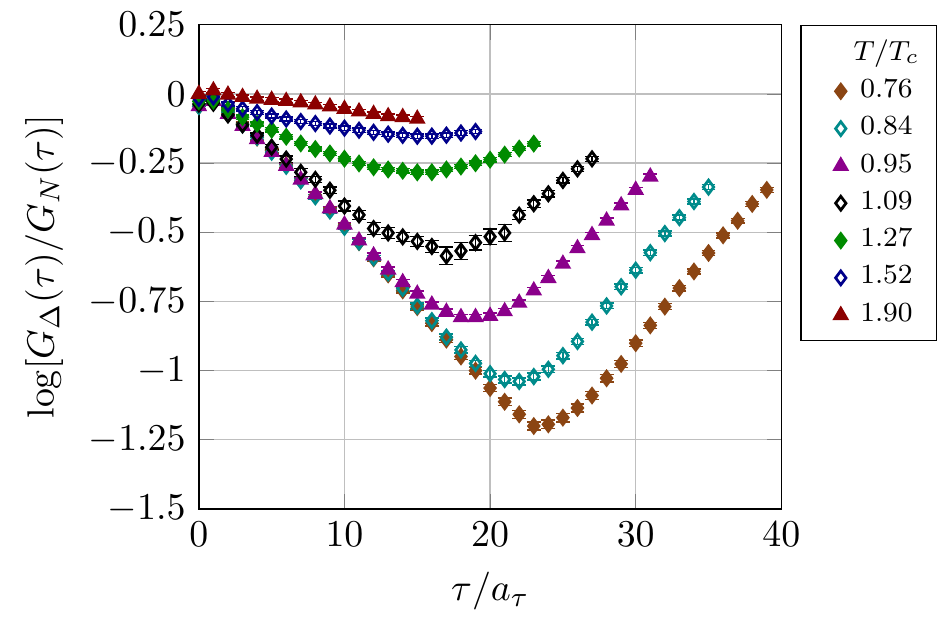} 
\caption{Logarithm of the ratio between the $\Delta$ and $N$ correlators, $G_\Delta(\tau)/G_N(\tau)$, for different temperatures, all normalised by $G_\Delta(0)/G_N(0)$ at $T/T_c=1.90$.
}
\label{fig:DN}
\end{figure}

The $N$ and the $\Delta$ baryon have the same quark content but different spin structure. 
In the confined phase this results in the mass splittings listed in Table \ref{tab:mass}. In the positive-parity channel the mass splitting is of the order of 300 MeV at all four temperatures, consistent with the PDG; in the negative-parity channel the mass difference is larger than in the PDG, but so is the uncertainty.
 In the deconfined phase, however, the quarks are quasi-free and the spin structure may become less important.\footnote{We thank Thomas Cohen for raising this question.} To investigate this, we show in Fig.\ \ref{fig:DN} the logarithm of the ratio of the $\Delta$ and $N$ correlators. All ratios are normalised by a single constant factor, $G_\Delta(0)/G_N(0)$ at $T/T_c=1.90$.
As expected, this ratio is falling exponentially below $T_c$, due to the (approximately constant) mass difference between the $N$ and the $\Delta$ baryons (in both parity channels). Above $T_c$, however, we observe a flattening of the ratio, approaching 1 at the highest temperature. We interpret this as an approximate degeneracy in the $N$ and $\Delta$ channels at very high temperature, which would be of interest to study further analytically. 
We also note the qualitative change in the ratio immediately at $T/T_c=1.09$, consistent with the observed changes in the spectrum.

\section{Thermal baryon spectral functions}
\label{sec:spec}

\subsection{Results}

The information in the thermal correlators discussed above is also present in the corresponding spectral functions, via relation (\ref{eq:Grho})
\be
\label{eq:spec}
 G_\pm(\tau)  =  \int_{-\infty}^{\infty} \frac{d\omega}{2\pi}\, K(\tau,\om) \rho_\pm(\omega),
\qquad\qquad
 K(\tau,\om)  = \frac{e^{-\om\tau}}{1+e^{-\om/T}}. 
\ee
As is well-known \cite{Asakawa:2000tr}, a simple inversion of this type of relation, using numerically determined correlators, is not possible. Hence we use the Maximum Entropy Method MEM \cite{Asakawa:2000tr,Bryan}, which extremises a combination of the standard likelihood ($\chi^2$) function, determined by the data, and an entropy function,
\be
 S = \int_{-\infty}^\infty \frac{d\om}{2\pi}\, \left[ \rho(\om) -m(\om)  - \rho(\om)\ln \frac{\rho(\om)}{m(\om)} \right],
 \ee
 encoding prior knowledge, via the default model $m(\om)$. The conditional probability to be extremised is of the form $\exp(-\half \chi^2+\alpha S)$, with $\alpha$ a parameter balancing the relative importance of the data and the prior knowledge. Both $m(\om)$ and $\alpha$ are further discussed below. 
 In the past 15 years, this method, and related ones, have been used by a number of groups, mostly for mesonic correlators, i.e.\ charmonium, the dilepton rate and the electrical conductivity, see e.g.\ Refs.\ \cite{Umeda:2002vr,Asakawa:2003re,Datta:2003ww,Aarts:2007pk,Aarts:2007wj,Ding:2010ga,Ohno:2011zc,Amato:2013naa,Borsanyi:2014vka,Aarts:2014nba,Brandt:2015aqk,Ding:2016hua}.
  Applications to bottomonium, in which some simplifications occur, can be found in Refs.\  \cite{Aarts:2011sm,Aarts:2013kaa,Aarts:2014cda,Kim:2014iga}. Here we give the first application to baryons.

Generic details of our implementation can be found in previous work \cite{Aarts:2007wj,Aarts:2011sm,Aarts:2014cda,Aarts:2014nba}. Here we briefly mention some differences with the bosonic (mesonic) case. We are interested in the spectrum for both positive and negative $\om$, since $\rho_-(\om)=-\rho_+(-\om)$. Hence the negative part of the spectrum of $\rho_+$ informs us of $\rho_-$, and vice versa. To bring the spectral relation (\ref{eq:spec}) to a numerically tractable form, we employ a cutoff $-\om_{\rm max} < \om < \om_{\rm max}$, with $a_\tau\om_{\rm max}= 3.0$ ($\om_{\rm max} =  16.9$ GeV). The remaining finite interval is discretised using $N_\om=2000$ bins. We have varied both $\om_{\rm max}$ and $N_\om$ to verify robustness. 
In the MEM analysis we used all the euclidean-time points, except for the time slices closest to the source and sink. At the lowest temperature, we have left out the points around the minimum of the correlators; this will be further discussed below.
As default model, we use a featureless constant, $m(\om)=m_0$, where the value of $m_0$ is determined by a fit to the correlation function using $\rho(\om)=m_0$ in Eq.\ (\ref{eq:spec}). 
Above $T_c$ we have fixed the default model to ensure a similar normalisation for all temperatures. 
We come back to the choice of default model below as well.

\begin{figure}[t]
\begin{center}
\includegraphics[width=0.48\textwidth]{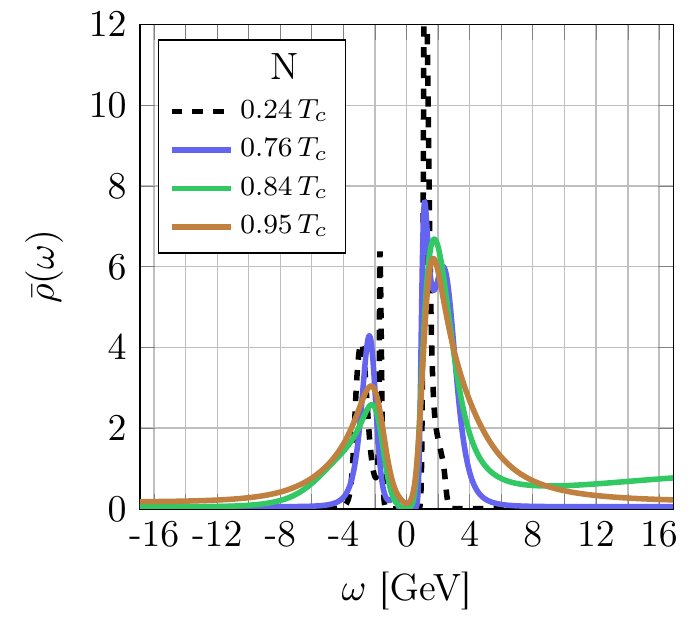}
\includegraphics[width=0.48\textwidth]{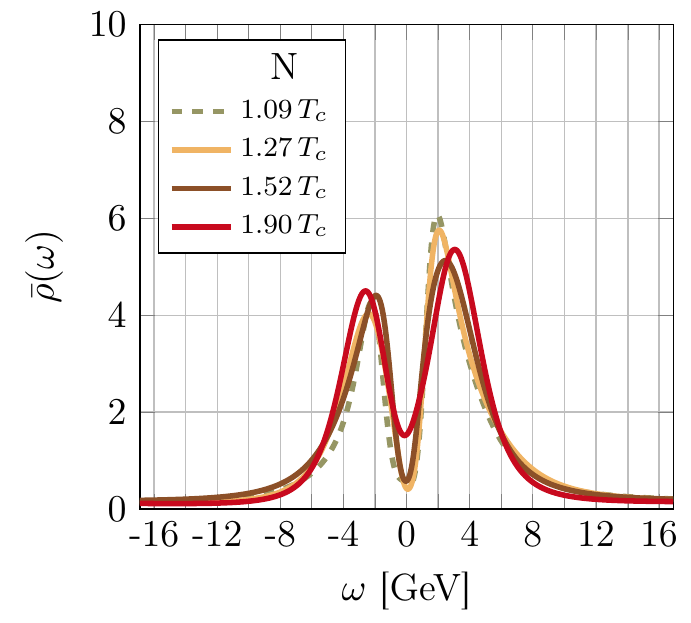}
\includegraphics[width=0.48\textwidth]{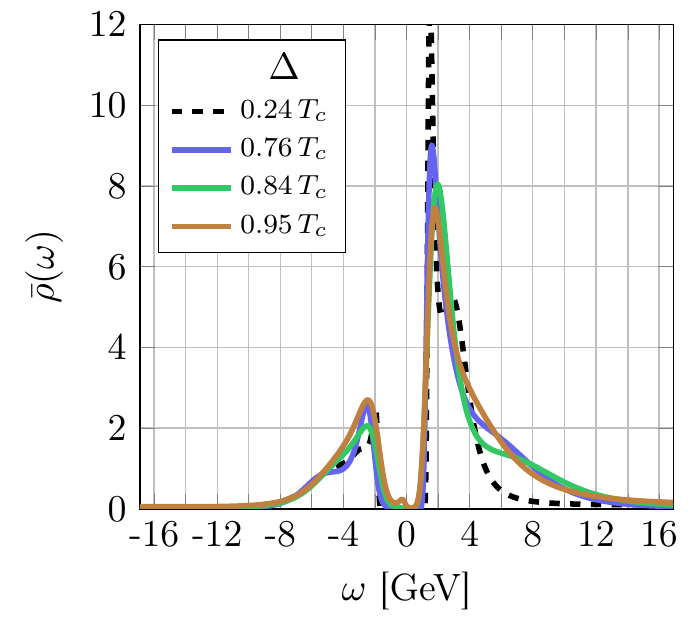}
\includegraphics[width=0.48\textwidth]{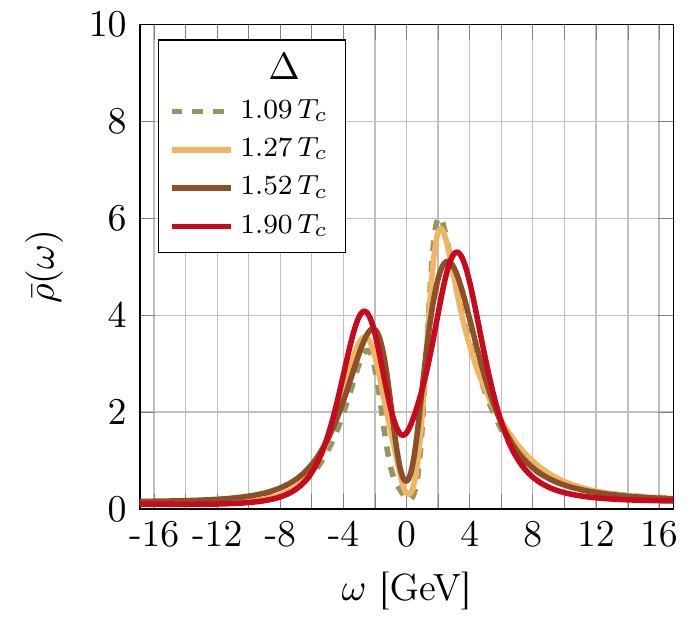}
\includegraphics[width=0.48\textwidth]{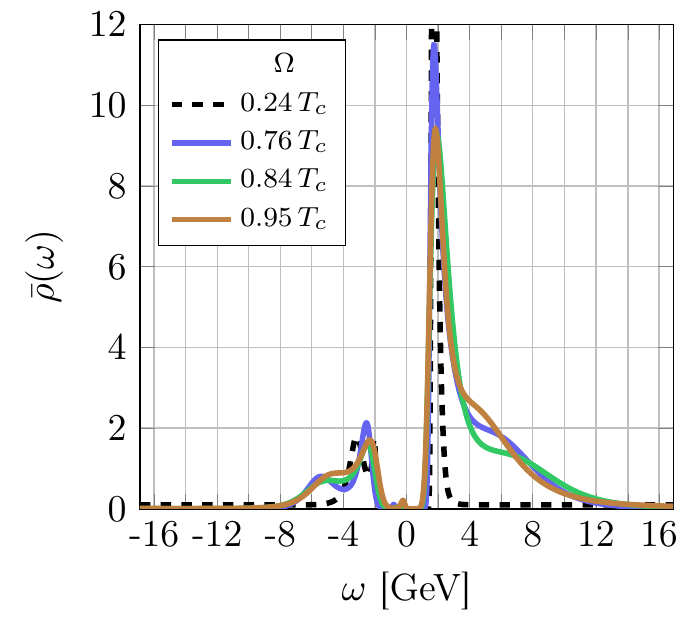}
\includegraphics[width=0.48\textwidth]{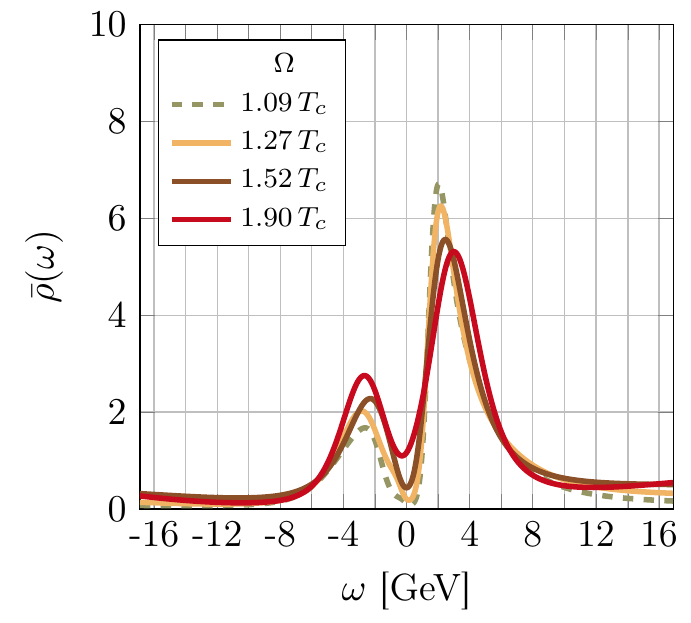}
\caption{Spectral functions below (left) and above (right) $T_c$ in the $N$ (top), $\Delta$ (centre) and $\Omega$  (bottom) channels.
}
\label{fig:3baryons}
\end{center}
\end{figure}

We now discuss the results. We have performed MEM on the normalised correlators $G_+(\tau)/(a_\tau G_+(\tau=0))$ and denote the associated dimensionless spectral functions with $\bar\rho(\omega)$. We note that the normalisation only affects the vertical scale but not the $\om$ dependence.
 Fig.~\ref{fig:3baryons} contains the spectral functions in the three channels below $T_c$ (left) and above $T_c$ (right).
 Spectral information for the positive-parity channel can be found at $\omega>0$, whereas $\omega<0$ refers to the negative-parity channel. Below $T_c$, the groundstate peaks on the positive-parity side are clearly visible and their positions agree with $m_+$, discussed in the previous section. Excited states are suppressed, due to the choice of smearing parameters.
  Some broadening is observed as the temperature is increased, but given the data and resolution, it is not clear whether this is a physical effect or due to the limitations of MEM. The negative-parity groundstates are visible as well, but are considerably less pronounced. The asymmetry between the positive- and negative-parity sides below $T_c$ is, however, clearly visible.

Above $T_c$, sharp groundstate peaks are no longer discernible. 
The broad peaks present above $T_c$ are most likely a combination of physical spectral features for deconfined quarks, as seen at very high temperature in perturbation theory \cite{Praki:2015yua,Praki:inprep}, and lattice artefacts due to the finite Brillouin zone, similar to in the mesonic case. To make this statement more quantitative would require a repetition of the calculation on finer lattices, which is one of our future aims.
Nevertheless, parity doubling manifests itself as $\rho_+(\om) = \rho_+(-\om)$, see Eq.\ (\ref{eq:rhopd}).
Hence the most important feature here is the emerging symmetry between the positive- and negative-parity sides as the temperature is increased. This is clearly visible for the $N$ and $\Delta$ channels, in which the position and height of the main features become comparable at positive and negative $\om$.
On the other hand, parity doubling is not yet complete in the $\Omega$ channel, as the positive-parity side is still enhanced. Nevertheless, the difference with spectral functions in the confined phase is manifest.
This is consistent with the analysis of the correlators above. 
 We note that in these plots we have not shown error bands for clarity; these will be discussed below.
 
\begin{figure}[t]
\centering
\includegraphics[width=0.50\textwidth]{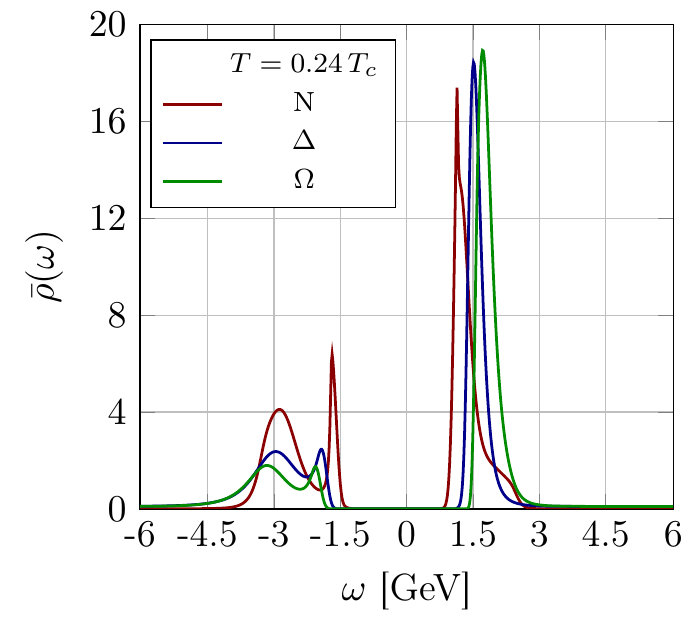}
\includegraphics[width=0.48\textwidth]{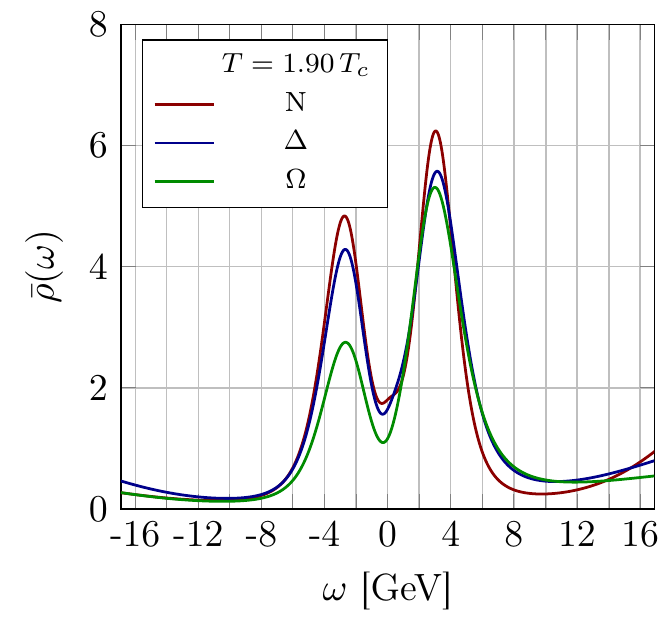}
\caption{Comparison between the $N$, $\Delta$ and $\Omega$ spectral functions at the lowest (left) and highest (right) temperatures.}
\label{fig:complowhigh}
\end{figure}

The combined results in all three channels are shown in Fig.~\ref{fig:complowhigh}, at the lowest (left) and highest (right) temperature. The difference between the spectral functions in the confined and deconfined phase is clear. We also note that below $T_c$ the negative-parity state is best visible in the nucleon sector.

\subsection{Default model and operator dependence}

\begin{figure}[t]
\begin{center}
 \includegraphics[width=0.50\textwidth]{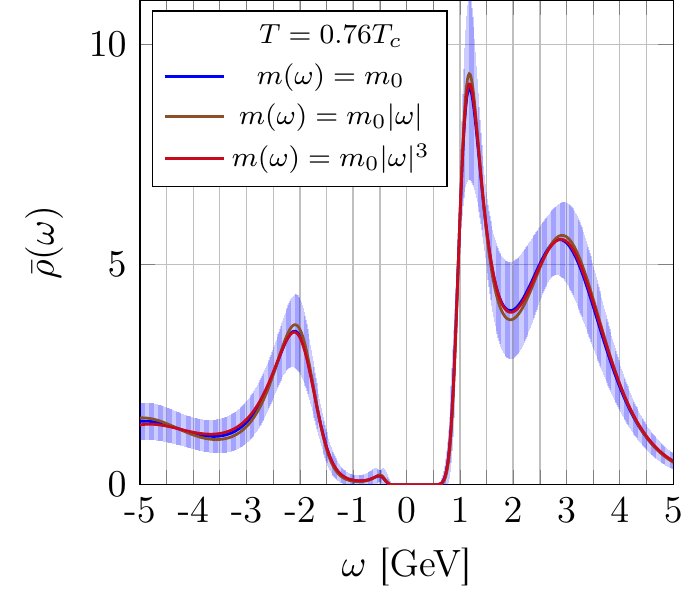}
 \includegraphics[width=0.49\textwidth]{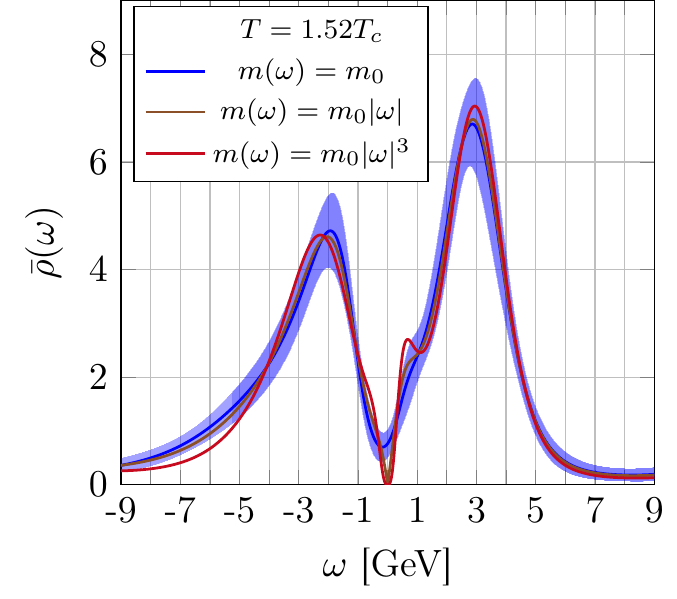}
\caption{Default model dependence in the nucleon channel at $T/T_c=0.76$ (left) and 1.52 (right).
 For clarity error bands are shown for one default model only.
 }
\label{fig:default}
\end{center}
\end{figure}

We now discuss some systematic effects in the construction of the spectral functions. We start with the default model dependence. The results above were obtained with a flat default model, $m(\om)=m_0$. We have also used $m(\om)=m_0|\om|$ and  $m(\om)=m_0|\om|^3$, where in each case $m_0$ is determined by a fit to the correlation function. The absolute value ensures positivity. In the continuum theory at leading order in weak coupling  \cite{Praki:2015yua,Praki:inprep}, the spectral functions increase as $|\om|^5$ for large $|\om| \gg T, m_q$, but this behaviour is modified on a finite lattice   \cite{Praki:2015yua,Praki:inprep}.
  Results are shown in Fig.\ \ref{fig:default}. The error band indicates the variation with the $\alpha$ parameter using Bryan's method \cite{Bryan} and is shown for one default model only, for clarity. We observe that  even though the default models are widely different, the resulting spectral functions are consistent within the uncertainty. 
 The second peaks in the confined phase at both $\om>0$ and $\om<0$ are presumably a combination of excited states and lattice artefacts. Whether a structure is due to a finite lattice cutoff or represents a physical feature can ultimately be tested by repeating the computation at smaller lattice spacings. One may also test the robustness with regard to the operators used, to which we turn now.
 
The dependence on the operator and the amount of smearing requires some  discussion. In previous studies in the mesonic sector, it has been  common to use a fixed local operator of the form $\bar\psi\Gamma\psi$, without smearing.\footnote{For charmonium, smearing has been employed in Ref.\ \cite{Umeda:2002vr}.} Locality is well motivated when the problem under investigation is related to a symmetry, such as electromagnetism (electrical conductivity, charge diffusion, dilepton production) and in Refs.\  \cite{Amato:2013naa,Aarts:2014nba} the conductivity and charge diffusion coefficient were determined  using the exactly conserved lattice vector current. For spectral questions at zero temperature, smearing and optimised operators aim to increase the overlap with the ground (or other) state, in such a way that the spectrum remains invariant, but spectral weight is redistributed. On the other hand, at finite temperature, where spectral functions are broadened and bound states eventually dissolve, spectral weight will potentially be nonzero at all energies. It is then less clear which features of the spectral function are invariant (and reflect the underlying physics) and which are e.g.\ operator dependent.

 \begin{figure}[t]
\begin{center}
 \includegraphics[width=0.49\textwidth]{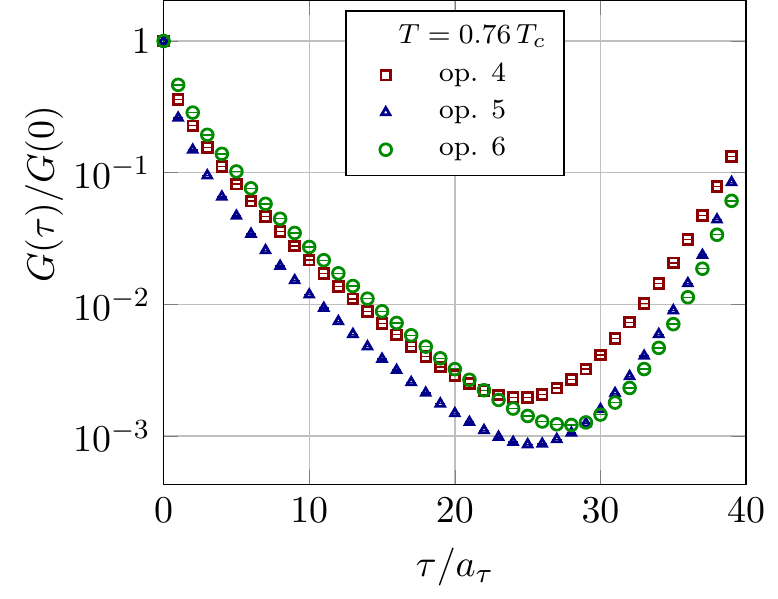}
 \includegraphics[width=0.49\textwidth]{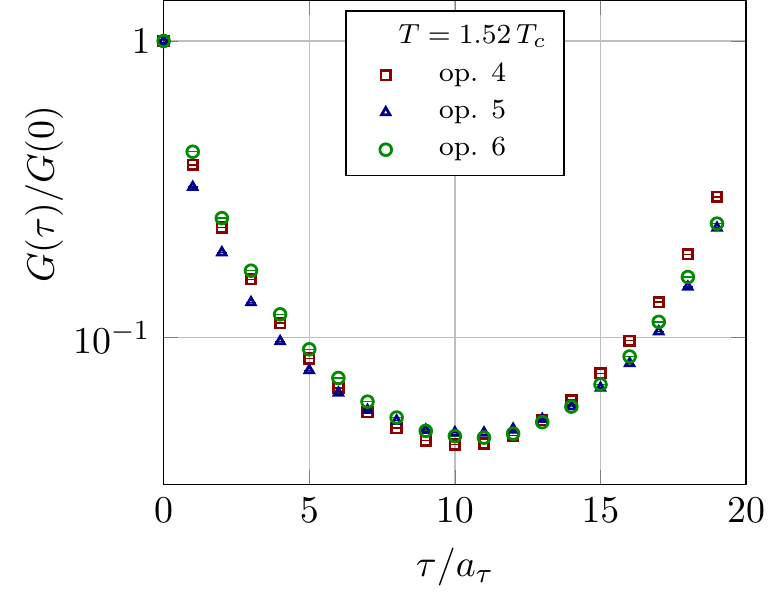}
 \includegraphics[width=0.50\textwidth]{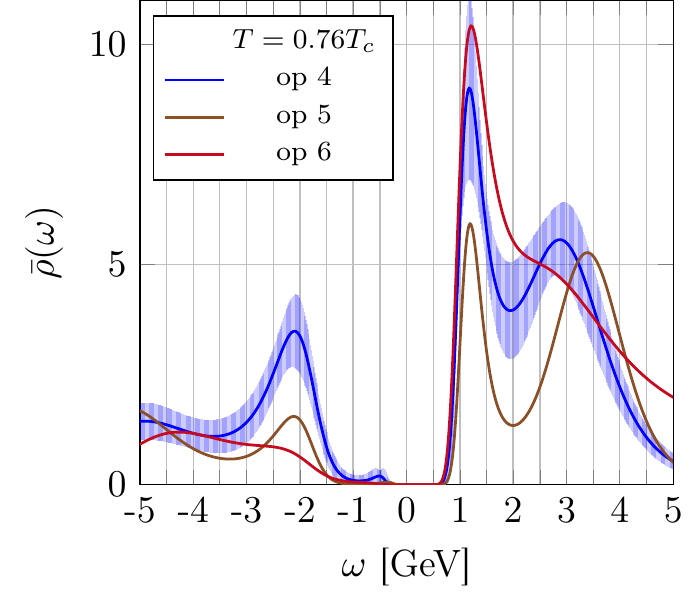}
 \includegraphics[width=0.49\textwidth]{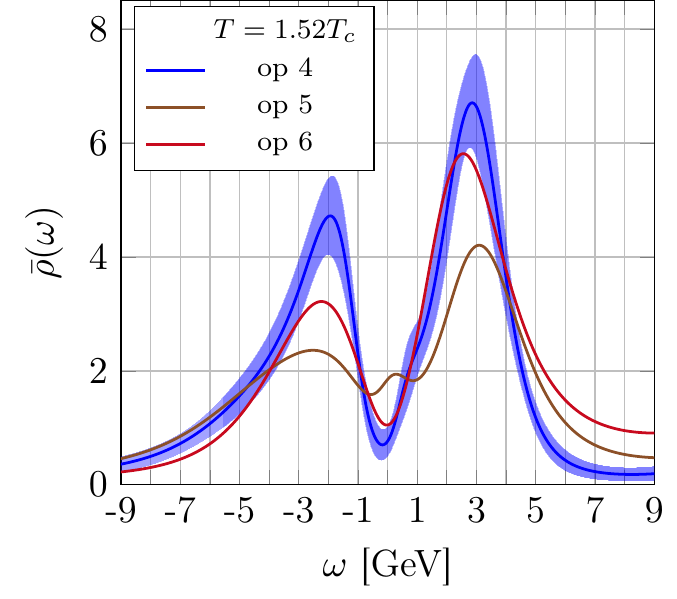}
\caption{Operator dependence in the nucleon channel at $T/T_c=0.76$ (left) and 1.52 (right), in the correlators (above) and the corresponding spectral functions (below).  For clarity, error bands are shown for operator 4 only.
}
\label{fig:operator}
\end{center}
\end{figure}

Smearing was already discussed to some extent in Ref.\ \cite{Aarts:2015mma}. Here we study the role of different operators.
We focus on the nucleon, with the interpolator chosen to be 
\be
O_N^{\alpha}(x) =\epsilon_{abc}  u_a^{\alpha}(x) \left(d^{^T}_b(x) C Y_n u_c(x)\right),
\ee
where $n=4,5,6$ and operators $Y_4=\gamma_5$, $Y_5=\gamma_4\gamma_5$ and $Y_6=\half (Y_4+Y_5)$ (this nomenclature follows Chroma  \cite{Edwards:2004sx}). Note that in the main part of the paper we have used operator $Y_4$.  The operator dependence is shown in Fig.\ \ref{fig:operator} for two temperatures. We observe that the correlators depend on the operator, as expected, since the overlap with ground- and excited states will differ. This manifests itself e.g.\ in the skewness of the correlator below $T_c$, while at high temperature approximate parity doubling is visible for all three operators.
Below $T_c$, we can quantify the spectral properties more precisely by comparing the masses $m_\pm^N$ from exponential fits, see Table \ref{tab:mass-op}.
We observe that the positive-parity mass $m_+^N$ is stable and consistent within the error. The negative-parity mass $m_-^N$ is consistent for operator 4 and 5, while for operator 6 the error is twice as large. This can be explained by noting that in Fig.\ \ref{fig:operator} (top, left) the correlator is most skewed for operator 6, which leads to the smallest temporal  range available on the negative-parity side, which is then reflected in the larger uncertainty.

Fig.\ \ref{fig:operator} (bottom) shows the corresponding spectral functions,  where at $T/T_c=0.76$ we observe groundstate peaks on the positive-parity side for all three operators. 
The position of the second peak  at $\om>0$ depends on the operator used; hence no physical relevance can be assigned to it.
On the negative-parity side the overlap with the groundstate is less pronounced. In particular operator 6 seems to have especially poor overlap with low-energy features on the negative-parity side.
 Just as above, this finding can be understood from the asymmetric shape of the correlator: the number of data points available for MEM is very limited. 
 At $T/T_c=1.52$, the approximate symmetry between the two parity sides is emerging, with the positive side still slightly enhanced, for all three operators. The fact that the overall area under the spectral curves appears different is related to the choice of normalisation. Yet the emerging symmetry, i.e.\ parity doubling, is present in all three cases, independent of the operator.

\begin{table}[t]
\centering
\begin{tabular}{ccccc}
$T/T_c$ & operator & 4 & 5 & 6  \\
\midrule
0.24 & $m^N_+$ [MeV]  & 1157(13) & 1156(13) & 1156(13) \\
	& $m^N_-$ [MeV]  & 1779(52) & 1824(48) & 1934(101) \\
\midrule
0.76 & $m^N_+$ [MeV]  & 1192(39)  &  1190(45)  & 1212(47) \\
	& $m^N_-$ [MeV]  & 1628(104) & 1698(106) & 1548(201) \\
\midrule
\end{tabular}
\caption{Operator dependence of $m_\pm^N$ at $T/T_c=0.24, 0.76$. }
\label{tab:mass-op}
\end{table}

\begin{figure}[t]
\centering
\includegraphics[width=0.48\textwidth]{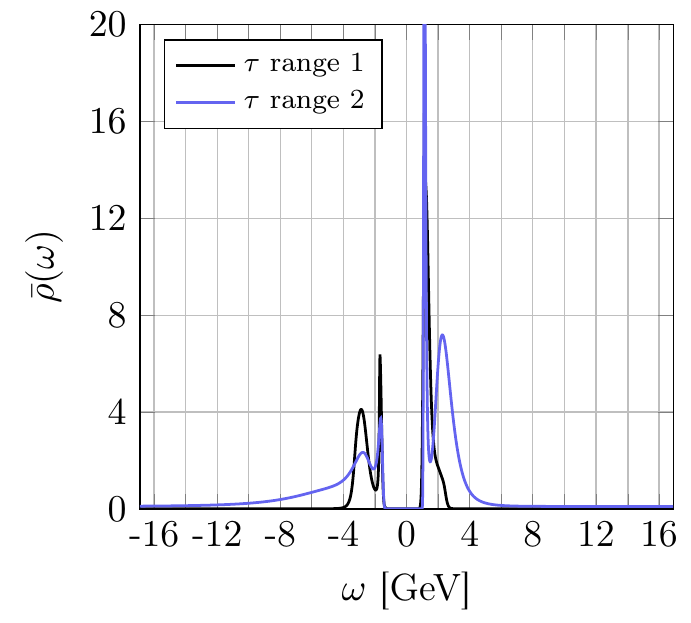}
\includegraphics[width=0.47\textwidth]{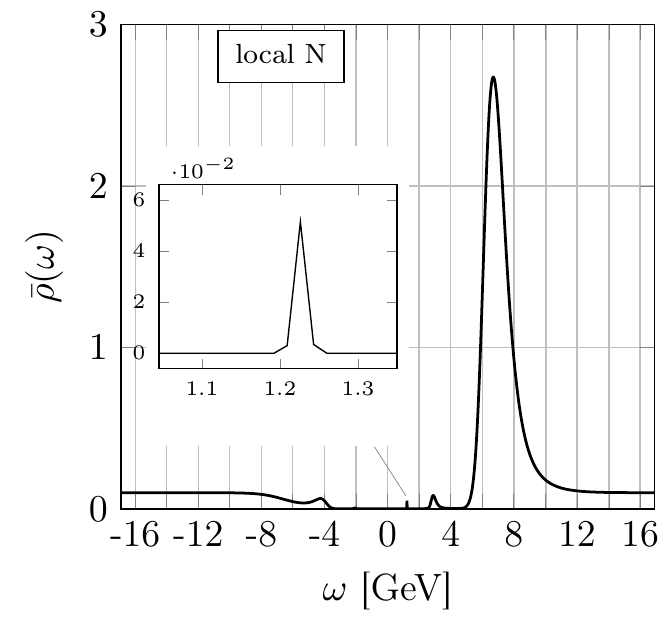} 
\caption{
Nucleon spectral function at $T/T_c = 0.24$. 
Left: Dependence on the time range used in the MEM analysis;  $\tau/a_\tau=[20,34] \cup [106,119]$ for range 1 and  $\tau/a_\tau=[6,66] \cup [90,126]$ for range 2.
Right:  Result obtained without smearing, using operator 4 as local sources and sinks. The inset shows a blow-up around the positive-parity groundstate.
}
\label{fig:spec-zoom}
\end{figure}

At the lowest temperature, we left out the points around the minimum of the correlators, both in the mass fits and the spectral function analysis, to handle a (mild) signal-to-noise problem. The effect of choosing various time ranges in the MEM analysis is shown in Fig.\ \ref{fig:spec-zoom} (left). For both ranges the groundstates are clearly distinguishable and in agreement, while differences appear for the possible excited states, which is as expected.

In the results presented above, smearing was used to single out the groundstate at low temperature and suppress contributions from highly excited states at all temperatures. As a final result we show in Fig.\ \ref{fig:spec-zoom} (right) the spectral function obtained at the lowest temperature in the nucleon channel, using local sources and sinks, i.e.\ without smearing. We observe a large contribution at higher energy, which is however not related to the low-energy states discussed above. The groundstate in the positive-parity channel is in fact still visible, as indicated in the inset, albeit much suppressed. When taken at face value, the mass is larger than found above, which is presumably due to the difficulty of extracting a signal from the local correlator. 
This figure therefore indicates the importance of smearing in this analysis, from a spectral function point of view.

In conclusion, we find that smearing and the choice of operator affects the correlators and hence the associated spectral functions at all temperatures. This is expected. At zero temperature, the masses of the groundstates are stable against these variations, as long as the groundstates are clearly identifiable. At nonzero temperature, the information gleaned from spectral functions is at a more qualitative level. Nevertheless, the conclusions drawn from the correlators and spectral functions are in agreement.


\section{Conclusion}

We studied the fate of the $N$, $\Delta$ and $\Omega$ baryons as the temperature is increased, using simulations with $N_f=2+1$ flavours of light quarks on anisotropic lattices. In the hadronic phase, we observed a strong temperature dependence of the groundstate masses for the negative-parity baryons, while the masses of the positive-parity baryons are stable up to the deconfinement transition. The temperature dependence is such that the positive- and negative-parity groundstates become approximately degenerate close to this transition. Degeneracy, i.e.\ parity doubling, is expected to coincide with chiral symmetry restoration and hence the transition from the hadronic to the quark-gluon plasma, but the precise manner in which this occurs is not known a priori. It would therefore be interesting to compare and contrast our nonperturbative predictions with model approaches, such as those discussed in Refs.\ \cite{Detar:1988kn,Nemoto:1998um,Jido:1998av,Zschiesche:2006zj,Steinheimer:2011ea,Benic:2015pia,Motohiro:2015taa,Nishihara:2015fka,Hohler:2015iba}, to reach further insight and understanding.

In the deconfined phase, we found strong indications that the light baryons no longer exist. Here we study parity doubling directly from an analysis of the correlators, using the $R$ parameter (\ref{eq:ratioR}), relating the positive- and negative-parity channels. We find a clear signal for the emergence of parity doubling, with the $R$ parameter acting as a quasi-order parameter. In the case of the $\Omega$ baryon, with the heavier $s$ quark, we find that parity doubling is not yet fully realised for the temperatures we considered. The effect of the quark mass is expected to vanish at higher temperatures, as $m_q/T\to 0$.

The conclusions from the correlator analysis are supported by the results obtained from the associated spectral functions. In the baryonic sector in vacuo, it is well understood that smearing and the use of optimised operators are essential to find clear signals for the ground- and other states. At finite temperature, with nonzero spectral weight at all energies, it is not immediately clear how to proceed with smearing and operator choice.
 In this paper we choose to optimise the smearing parameters and operators at zero temperature and keep them fixed as the temperature increases. With this prescription we found it is possible to obtain quantitative results from the correlator analysis and qualitative insight from the spectral functions, which are mutually consistent. It would be interesting to consider this question further and e.g.\ employ variational bases, widely used in vacuum, also at finite temperature, as suggested in Ref.\ \cite{Harris:2016usb}. 

As an outlook, there are various directions in which this study can be taken further, in addition to those mentioned above. From the viewpoint of lattice QCD, an important role is played by chiral symmetry. Since the Wilson-clover quarks employed here break chiral symmetry at short distances (and the two light flavours are still somewhat heavy), it would be interesting to repeat this calculation with manifestly chiral (domain wall/overlap) fermions. The signal for parity doubling should then be easily visible in the correlators, without the need to suppress short-distance contributions. A physical question is related to the role of strangeness, since a finite $s$ quark mass breaks chiral symmetry explicitly.  For the $\Omega$ baryon, we indeed observed the effect of the strange quark mass in the signal for parity doubling, but a more comprehensive study of strange baryons would enlighten this further. Finally, we observed strong in-medium effects for the negative-parity baryons in the hadronic phase. 
It would hence be interesting to investigate whether and how this affects heavy-ion phenomenology, e.g.\ in the context of the hadron resonance gas or the statistical hadronisation model \cite{Stachel:2013zma}.


\section*{Acknowledgments}

We thank Sin\'ead Ryan, Tim Burns, Thomas Cohen, Claudia Ratti and Rene Bellwied  for discussion.
The work has been supported by STFC grant ST/L000369/1, SNF grant 200020-162515, ICHEC, the Royal Society, the Wolfson Foundation and the Leverhulme Trust, and has been performed in the framework of COST Action CA15213 THOR.
We are grateful for the computing resources
made available by HPC Wales. This work used the DiRAC Blue Gene Q Shared
Petaflop system at the University of Edinburgh, operated by the Edinburgh
Parallel Computing Centre on behalf of the STFC DiRAC HPC Facility
(www.dirac.ac.uk). This equipment was funded by BIS National E-infrastructure
capital grant ST/K000411/1, STFC capital grant ST/H008845/1, and STFC DiRAC
Operations grants ST/K005804/1 and ST/K005790/1. DiRAC is part of the National
E-Infrastructure.


\providecommand{\href}[2]{#2}\begingroup\raggedright\endgroup

\end{document}